\documentclass[aps,prb,groupedaddress,superscriptaddress,showpacs]{revtex4}
\usepackage{graphicx}          % Include figure files
\usepackage{dcolumn}           % Align table columns on decimal point
\usepackage{bm}                % bold math
\usepackage[usenames]{color}
\usepackage[normalem]{ulem}
\usepackage{amsmath}
\usepackage{amssymb}

\begin{document}

\title{Modeling Near-Surface Bound
Electron States in Three-Dimensional Topological Insulator: Analytical
and Numerical Approaches}

\author{V.N. Men'shov}
 \affiliation{NRC Kurchatov Institute, Kurchatov Sqr. 1, 123182 Moscow, Russia}
\affiliation{Donostia International Physics Center (DIPC),
             20018 San Sebasti\'an/Donostia, Basque Country,
             Spain\\}

\author{V.V. Tugushev}
 \affiliation{NRC Kurchatov Institute, Kurchatov Sqr. 1, 123182 Moscow, Russia}
 \affiliation{A.M. Prokhorov General Physics Institute, Vavilov str. 38, 119991
Moscow, Russia} \affiliation{Donostia International Physics Center
(DIPC), 20018 San Sebasti\'an/Donostia, Basque Country,
             Spain\\}

\author{T.V. Menshchikova}
 \affiliation{Tomsk State University, 634050 Tomsk, Russia}
\affiliation{Donostia International Physics Center (DIPC),
             20018 San Sebasti\'an/Donostia, Basque Country,
             Spain\\}

\author{S.V. Eremeev}
 \affiliation{Institute of Strength Physics and Materials Science,
634021, Tomsk, Russia}
 \affiliation{Tomsk State University, 634050 Tomsk, Russia}
\affiliation{Donostia International Physics Center (DIPC),
             20018 San Sebasti\'an/Donostia, Basque Country,
             Spain\\}

\author{P.~M. Echenique}
 \affiliation{Donostia International Physics Center (DIPC), 20018 San Sebasti\'an/Donostia, Basque Country, Spain\\}
\affiliation{Departamento de F\'{\i}sica de Materiales UPV/EHU,
Centro de F\'{\i}sica de Materiales CFM - MPC and Centro Mixto
CSIC-UPV/EHU, 20080 San Sebasti\'an/Donostia, Basque Country, Spain}

\author{E.V. Chulkov}
\affiliation{Donostia International Physics Center (DIPC),
             20018 San Sebasti\'an/Donostia, Basque Country,
             Spain\\}
\affiliation{Departamento de F\'{\i}sica de Materiales UPV/EHU,
Centro de F\'{\i}sica de Materiales CFM - MPC and Centro Mixto
CSIC-UPV/EHU, 20080 San Sebasti\'an/Donostia, Basque Country, Spain}

%\ead{vnmenshov@mail.ru}

\date{\today}

\begin{abstract}
We apply both analytical and ab-initio methods to explore
heterostructures composed of a three-dimensional topological
insulator (3D TI) and an ultrathin normal insulator (NI) overlayer
as a proof ground for the principles of the topological phase
engineering. Using the continual model of a semi-infinite 3D TI we
study the surface potential (SP) effect caused by an attached
ultrathin layer of 3D NI on the formation of topological bound
states at the interface. The results reveal that spatial profile and
spectrum of these near-surface states strongly depend on both the
sign and strength of the SP. Using ab-initio band structure
calculations to take materials specificity into account, we
investigate the NI/TI heterostructures formed by a single
tetradymite-type quintuple or septuple layer block and the 3D TI
substrate. The analytical continuum theory results relate the
near-surface state evolution with the SP variation and are in good
qualitative agreement with those obtained from density-functional
theory (DFT) calculations. We predict also the appearance of the
quasi-topological bound state on the 3D NI surface caused by a local
band gap inversion induced by an overlayer.
\end{abstract}

\pacs{73.20.-r, 75.70.Cn}

%\submitto{\NJP}

\maketitle

\section{Introduction}

It is generally recognized that there is one-to-one correspondence
between the presence of nontrivial topological invariants
characterizing the bulk electron states of a crystal and the
appearance of specific electron modes localized at the crystal
boundary \cite{Fu2,Moore1,Essin,Hasan,Qi,Okuda,Ando2013}. This
statement, which is known as the bulk-boundary correspondence
theorem, reflects profound interrelation between the  interior and
exterior electron states of the truncated crystal. While the
formulated assertion is based on general arguments of the
topological concept for solids, in real materials and/or
heterostructures, the effect of the peculiar bulk properties on the
surface may be quite intricate. This problem has been widely
discussed in the context of an existence of topological bound states
on the surface of a three-dimensional topological insulator (3D TI)
or at the interface between 3D TI and topologically trivial material
in various hybrid structures\cite{Hasan,Qi,Okuda,Ando2013}.
Angle-resolved photoemission spectra have given evidence for the
Dirac-like dispersion and the momentum-dependent spin texture of the
3D TI surface states in a family of Bi- and Sb-based narrow-gap
semiconductors with strong spin-orbit coupling (SOC)
\cite{Hasan,Qi,Okuda,Eremeev1}, which leads to the inverted band
structure characterized by nontrivial topological invariants.

Although the topologically protected surface states are often
considered as the most important and even decisive property of 3D
TIs, in practice, the specific manifestations of the
boundary-related electron properties in semiconductor materials with
an inverted energy gap go far beyond the bulk-boundary
correspondence paradigm. In other words, being formally correct, the
topological arguments tell us too little about characteristics of
the topologically protected surface/interface states (e.g., the
details of the dispersion, actual length scale, spin texture) or
what other in-gap electron states might form at the real boundary of
a 3D TI material. While a rapid progress has been made in
investigations of the vacuum-terminated TI surfaces
\cite{Hasan,Qi,Okuda,Eremeev1}, it is still challenging to control
the properties of the Dirac states under the surface modification in
complex situations when a 3D TI is brought into the contact with
some substances. Recent experiments on 3D TIs have demonstrated that
these states are notably sensitive to external perturbations, such
as chemical doping of the surface via deposition of both magnetic
and nonmagnetic elements \cite{Wray,Scholz,Valla}, oxidation of
air-exposed samples \cite{Kong}, change of the surface termination
\cite{Eremeev1,Miao}, capping layers and interfaces with other
materials \cite{Jenkins,Berntsen}, applying an external gate voltage
\cite{Chen,Checkelsky}, etc.

Perturbation of a bulk crystal potential exists naturally in
truncated crystals and, in particular, 3D TIs, and creates a surface
potential (SP) affecting electron properties of a crystal near/at
the surface. In Refs.~\cite{Bianchi,Bahramy} it was argued that a
bulk-truncated surface of bismuth-chalcogenides can develop complex
electronic structure in which the Dirac states coexist with the
conventional states of the two-dimensional electron gas (2DEG) in
the quantum well appearing near the 3D TI surface due to the
band-bending effect. The \emph{ab initio}
calculations~\cite{Menshch2011,Eremeev2} have shown that an
expansion of the van-der-Waals (vdW) spacing in layered 3D TIs
caused by intercalation of deposited atoms leads to a simultaneous
emergence of 2DEG bands localized in the subsurface region.
Moreover, the expansion of the vdW spacing also leads to a
relocation of the Dirac topological states to the lower quintuple
layers~\cite{Eremeev2}. Wang~\emph{et al} have studied the effects
of surface modification on the topological surface state in
Bi$_2$Se$_3$ using first-principles calculations and shown that
Bi-capping and Se-removing can move the Dirac point upwards and slow
flatten the topological surface bands~\cite{Wang}. The short-range
chemical forces related to dangling-bonds on the surface of a
thallium-based ternary chalcogenides TIs (TlBiTe$_2$, TlBiSe$_2$,
TlSbTe$_2$, and TlSbSe$_2$) produce strong surface states
\cite{Eremeev3} which can be removed by thallium
adatoms~\cite{Kuroda2013}. In Ref.~\cite{LZhao} it was suggested
that the Dirac point of the helical surface states can be
significantly shifted by applying uniaxial strain.

Many exciting physical properties of the Dirac helical
quasiparticles (in particular, spin-dependent transport) are
predicted to provide good opportunities for different spintronic
applications \cite{Garate,Yu,Fujita,Pesin}. To fully embody these
promising ideas in devices, one requires multiple interfaces with
the topologically trivial materials rather than a single pristine
surface of TI. Using density functional theory to design
superlattice structures based on Bi$_2$Se$_3$, it was shown that an
interface state with an ideal Dirac cone is caused by alternating
the layers of 3D TI and 3D normal insulator (NI) \cite{Song}. The
authors of Ref.~\cite{Zhang} have studied theoretically the
Sb$_2$Se$_3$/Bi$_2$Se$_3$ heterostructures, the constituents of
which possess the Bloch functions of the same symmetry. They found
that the probability maximum of the Dirac state largely moves from
the topologically nontrivial Bi$_2$Se$_3$ into the region of the
topologically trivial Sb$_2$Se$_3$. On the other hand, ARPES
experiments~\cite{YZhao} provide the direct evidence that the
surface state of top surface of the heterostructure containing
single quintuple layer (QL) of Bi$_2$Se$_3$ on 19QLs of Bi$_2$Te$_3$
is similar to the surface state of Bi$_2$Se$_3$. Moreover, the
transport measurements~\cite{YZhao} show that the studied
heterostructure behaves more like Bi$_2$Se$_3$ even though there is
only 1QL Bi$_2$Se$_3$ layer grown on 19QLs Bi$_2$Te$_3$. In
Refs.~\cite{Wu,Li} it was established that, as a result of
depositing a 3D NI overlayer (conventional semiconductor ZnM, M=S,
Se, and Te) onto the 3D TI substrate (Bi$_2$Se$_3$ or Bi$_2$Te$_3$),
the topological states can float to the top of the NI film, or stay
put at the NI/TI interface, or are pushed down deeper into 3D TI.
Recently Berntsen and colleagues have directly observed the Dirac
states at the Bi$_2$Se$_3$/Si(111) buried interface~\cite{Berntsen}.
Another photoemission study in Ref.~\cite{Nakayama} revealed an
existence of the interface topological states in the layered bulk
crystal (PbSe)$_5$(Bi$_2$Se$_3$)$_{3m}$, which forms a natural
multilayer heterostructure composed of TI and NI. The evidence of a
large shift of the Dirac point towards the conduction band edge
relative to the case of the 3D TI/vacuum interface, due to the
In$_2$Se$_3$ \cite{Jenkins} or Sb$_2$Se$_2$Te \cite{Tania} capping
layer on the epitaxial Bi$_2$Se$_3$ thin film, was reported
demonstrating a possibility of controlling the Dirac cone in 3D
TI-based systems.

Thus, the experimental and theoretical data exhibit that the real 3D
TI surface and 3D TI/NI interface possess very rich and diverse
physics, in particular, they can hold both topological and
non-topological (ordinary) in-gap states. It is well known that the
non-topological bound states can be created or deleted or altered,
depending on the both the sign and strength of SP, when 3D NI is
exposed to ambient conditions or put into the contact with other
materials. On the contrary, in 3D TI, the topological order itself
is robust against such the influences so that it can be completely
destroyed only under the drastic perturbation~\cite{Chena,Eremeev4}.
Nevertheless, the parameters of the topological states can undergo
remarkable changes with even moderate external perturbations.
Combination of the robustness of the topological states at the TI/NI
interfaces with the tunability of their parameters to the external
influence favours the design of the 3D TI/NI layered systems
possessing suitable band structure, charge distribution and spin
texture. The efficient design of the 3D TI/NI systems can be
realized by combining analytic and numerical methods.

In the present work, we consider a special type of the 3D TI/NI
heterostructures of particular interest, which contain an ultrathin
film of nonmagnetic NI (overlayer) artificially deposited on a
relatively thick film of 3D TI (substrate). Due to a specific
relation between electron affinities and band gap widths of the
substrate and overlayer materials, significant modifications of the
spectrum and wave function of the Dirac states are expected as
compared to the pristine 3D TI surface (i.e. the 3D TI/vacuum
interface), thus making it possible to obtain the 3D TI-based
heterostructure with tailor-made electron properties. One assumes
that the substrate film thickness is large enough to avoid sizable
hybridization of the bound states appearing at the opposite
boundaries of the film. At the same time, the minimal thickness of
the overlayer is formally limited by the condition of an electron
motion quantization in the 3D NI material. Under these restrictions,
to describe analytically the electron bound states near the surface
of the truncated 3D TI covered by the 3D NI overlayer, the continual
approach involving the method of effective surface potential (SP)
was offered in Ref.~\cite{Men}. In what follows, we will use the
term 'near-surface state' (see also Ref.~\cite{Men}) for
identification of the in-gap electron state localized in the
subsurface region and driven by the overlayer-induced SP. Just
recently, in Ref.~\cite{Tania}, several preliminary results
concerning the near-surface states in the 3D TI/NI heterostructures
with realistic material parameters has been obtained within the
numerical simulations based on density functional theory (DFT).
Below , we  employ the two complementary approaches -- analytical
and numerical -- in order to elucidate thoroughly the important
question how a 3D NI overlayer affects the electron properties of 3D
TI/NI heterostructures.

In the framework of the analytical approach, it is instructive to
re-formulate this question in terms of the boundary conditions at
the TI/NI interface for the wave function of the system. It is clear
that, in systems composed of two topologically distinguishable
materials, the characteristics of the near-surface state depend
crucially on the choice of the boundary conditions, which still
remains highly disputable subject (for example, see
Refs.~\cite{Shan,Medhi,Michetti}). The wave function at the ideal
atomic interface between a pair of similar materials (e.g., 3D TI
Bi$_2$Se$_3$ and 3D NI Sb$_2$Te$_3$ have the same crystal symmetry)
satisfies the Ben-Daniel\&Duke boundary conditions~\cite{BenDaniel}.
While matching the wave function at the contact of two dissimilar
materials (e.g., such as Si and Bi$_2$Se$_3$) is complicated within
the $\mathbf{k}\mathbf{p}$ formalism because the envelope function
(EF) on each side of the interface are defined using distinct
orbital basis (see Ref.~\cite{Tokatly} and reference therein).
However, such complication proved to be circumvented for the
particular models describing different types of contacts within the
effective interface potential
concept~\cite{Men2010,Menshov,Men2013}.

Below, in the framework of the continual approach involving the SP
scheme, we formulate general boundary conditions for the long-range
envelope function of the truncated 3D TI and truncated 3D NI and
find the solution for the bound near-surface states. We restrict
ourselves to the situation when solely the orbital degree of freedom
of electrons is manipulated by external influence at the surface. We
succeed in general qualitative understanding of the dependence of
the energy spectrum and spatial profile of the near-surface states
on an effective SP. Furthermore, in order to elucidate the fine
details of the near-surface state transformation induced by the
overlayer, we employ the material-specific DFT calculations for the
3D TI/NI heterostructures. For the conceptual reasons, within an
effective SP scheme, we also discuss the near-surface states in the
fictitious heterostructures of other types, in which a substrate of
a 3D NI close to the quantum transition into a 3D TI phase is
covered with an overlayer of either NI or TI material.

The paper is organized as follows. In Sec.~2, we discuss the
effective SP concept, propose the model for a truncated TI covered
with an overlayer, and introduce the main ingredients and
assumptions of the problem within the continual approach. In Sec.~3,
for the case of a spin-independent surface perturbation caused by an
overlayer, we thoroughly investigate how the corresponding SP
modifies the electron energy spectrum and the EF spatial profile of
the Dirac-like near-surface state. In Sec.~4, we analyze main
features of the near-surface state in a situation when a  3D NI
substrate close to transition into a topological phase is covered
with a ultrathin overlayer of either a NI material or a TI one. To
corroborate the SP formalism results , in In Sec.~5, we apply the
DFT calculations and analyze the band structure and wave-function of
the bound near-surface states for a set of heterostructures formed
by a single tetradymite-type quintuple (QL) or septuple (SL) layer
blocks and 3D TI substrates. Finally, the main conclusions are
presented in Sec.~6.

\section{Surface potential concept and model hamiltonian}

Apart from the aforementioned simulations of the properties of 3D
TIs based on the first principle calculations, various continual
models have been discussed to describe relativistic fermions at the
TI boundary~\cite{Zhang,Liu,Shan}. There are theoretical studies of
the TI properties, which are routinely based on the simple
phenomenological 2D Hamiltonian for helical fermions with the linear
Dirac-cone-like energy-momentum dispersion under an external
influence~\cite{Fu}:
$\mathbb{H}_{s}=-iv(\mathbf{e}_z[\bm{\sigma}\times
\bm{\bigtriangledown}])+\mathbb{U}$, where $\mathbf{e}_z$ is the
unit vector normal to the surface, $v$ is the Fermi velocity,
$\bm{\sigma}$ is the vector composed of the Pauli matrices. It is
generally thought that a controllable external field, $\mathbb{U}$,
can be directly applied to the 3D TI surface to manage its electron
states. For instance, a spin-independent term $\mathbb{U}=I_{2\times
2}U$ could simulate the energy shift of the Dirac-cone point due to
an electrostatic potential caused by a nonmagnetic overlayer. In
turn, a spin-dependent term $\mathbb{U}\sim \sigma_{z}m$ could
generate a gapped spin-polarized surface state through an exchange
field proportional to the magnetization $\mathbf{m}=\mathbf{e}_z m$
applied along the normal to the surface of 3D TI which is in contact
with a ferromagnetic insulator~\cite{Garate,Tserkovnyak,Yokoyama}.
In this manner, to take into account a perturbation arising from the
external influence, the additional term $\mathbb{U}$ is simply
included in the 2D Hamiltonian $\mathbb{H}_{s}$, without a serious
analysis of the microscopic origin of both $\mathbb{H}_{s}$ and
$\mathbb{U}$. The vast majority of theoretical works restricts to
such the description and they predict many curious effects which can
be realized in the 3D TI-based structures. However, the 2D
Hamiltonian $\mathbb{H}_{s}$ can formally be derived from a relevant
3D Hamiltonian only under the free surface stipulation in the spirit
of Refs.~\cite{Zhang,Shan}. In the framework of the consistent
scheme, the bound near-surface states for the half-infinite 3D TI
are composed of the eigen-states of the corresponding bulk 3D TI
Hamiltonian. Meanwhile, note that so far nobody has written down the
full set of the orthogonal wave-functions (including the bound and
extended along $\mathbf{e}_z$-direction states) for the
$\mathbf{k}\mathbf{p}$ Hamiltonian of 3D TI in the half-infinite
geometry even under the free boundary conditions on the surface.
Strictly speaking, the bound states alone do not form the full basis
set suited for correct description of the $\mathbb{U}$ field effect
on 3D TI. The external surface perturbation excites electron density
in the bulk 3D TI region of a nanoscopic scale adjacent to the
surface. Both the bound and extended electron modes give rise to the
response of 3D TI to this perturbation. Upon placing the 3D NI on 3D
TI, besides the topological bound state, the so-called ordinary
bound state~\cite{Menshov} can arise near the interface due to the
hybridization between the NI and TI atomic orbitals through the
interface. Hence the near surface electron density perturbation of
the 3D TI has very complicated spatial, orbital and spin
configuration. The phenomenological 2D Hamiltonian $\mathbb{H}_{s}$
hardly could serve as a starting point for the correct analysis of
the configuration-dependent response of the 3D TI. So, to correctly
take into account the effect of an external perturbation on the
surface/interface electron states in the 3D TI, one has to directly
include the field $\mathbb{U}$ into the "true" 3D Hamiltonian of the
system.

A basic idea to go beyond the scope of the 2D model is the use of
the well-known $\mathbf{k}\mathbf{p}$ method~\cite{Bir}. To
characterize the band electron states of a bulk semiconductor,
$|n\mathbf{k}\rangle$ ($\mathbf{k}$ is a wave vector, $n$ is a band
index), in the region of the Brillouin zone around the point of band
extrema $\mathbf{k}_{0}$, the $\mathbf{k}\cdot\mathbf{p}$ method is
reputed to be accurate enough. Under a perturbation smooth on the
atomic scale, this method makes it possible to predict evolution of
the electron state wave function $\Psi_{n}(\mathbf{r})$ in terms of
a product of a slowly varying envelope function (EF)
$\theta_{n}(\mathbf{r})$ and the Bloch function of the unperturbed
crystal
$|n\mathbf{k}_{0}\rangle=\exp(i\mathbf{k}_{0}\mathbf{r})u_{n\mathbf{k}_{0}}(\mathbf{r})$
at the point $\mathbf{k}_{0}$:
$\Psi_{n}(\mathbf{r})=\theta_{n}(\mathbf{r})|n\mathbf{k}_{0}\rangle$,
$u_{n\mathbf{k}_{0}}(\mathbf{r})$ is the lattice periodic function.
The EF concept may also be applied to the description of localized
and resonant interface states in the semiconductor junctions of
different types. However, a relevant choice of the boundary
conditions for the function $\theta_{n}(\mathbf{r})$ remains an
unsettled question in this concept, in particular for the TI based
structures. The authors of Refs.~\cite{Zhang,Shan} impose the
so-called 'open' boundary conditions fixing all EF components to
zero at the crystal surface. This restriction formally simulates the
effect of vanishing of the quasiparticle wave function on the
infinitely high SP barrier. Nevertheless, it should be pointed out
that the zero constraint is not unique, and other options for
boundary conditions have been advocated in the literature. For
instance, in Ref.~\cite{Medhi} the problem is formulated in terms of
an energy functional whose minimization yields the so-called
'natural' boundary conditions, intermixing the magnitudes and
derivatives of different EF components. Both mentioned types of the
EF boundary conditions are extremely idealized and cannot adequately
take into consideration a sensitivity of the 3D TI electron states
to the surface modifications.

In this work we propose a formalism to directly incorporate the
surface perturbation effect into the 3D TI Hamiltonian. We derive
the appropriate EF boundary conditions through the construction of
the effective semi-phenomenological SP localized at the 3D TI
surface. The orbital and spin structure of the SP mimics induced
fields resulting from a surface perturbation. As shown below, the
structure and strength of the SP determine both the spatial and
spectral features of the topological states.

The low energy and long wavelength bulk electron states of the
prototypical TI, narrow-gap semiconductor of Bi$_2$Se$_3$-type, are
described by the four bands $\mathbf{kp}$ Hamiltonian with strong
SOC proposed in Refs.~\cite{Zhang,Liu}. Without a loss of
generality, we make use of the simple version of this Hamiltonian
in the form:
\begin{equation}\label{H-TI}
\mathbb{H}(\mathbf{k})=\Xi(\mathbf{k})\tau_{z}\otimes\sigma_{0}
+\mathrm{A}\tau_{x}\otimes(\bm{\sigma}\cdot\mathbf{k}),
\end{equation}
where $\Xi(\mathbf{k})=\Xi-\mathrm{B}k^{2}$, $\mathbf{k}$ is the
wave vector, $k=|\mathbf{k}|$, $\sigma_{\alpha}$ and $\tau_{\alpha}$
($\alpha=0,x,y,z$) denote the Pauli matrices in the spin and orbital
space, respectively. The Hamiltonian is written in the basis
$u_{\mathbf{k}_{0}}
=\{|+\uparrow\rangle,|-\uparrow\rangle,|+\downarrow\rangle,|-\downarrow\rangle\}$
of the four states at the $\Gamma$ point of the Brillouin zone with
$\mathbf{k}_{0}=0$. The superscripts $\pm$ denote the even and odd
parity states and the arrows $\uparrow\downarrow$ indicate the spin
projection onto the $z$ quantization axis. The Hamiltonian
(\ref{H-TI}) captures the remarkable feature of the band structure:
under the condition $\Xi\mathrm{B}>0$, the inverted order of the
energy terms $|+\uparrow(\downarrow)\rangle$ and
$|-\uparrow(\downarrow)\rangle$ around $\mathbf{k}_{0}=0$, which
correctly characterizes the topological nature of the system due to
strong SOC. The Hamiltonian (\ref{H-TI}) is particle-hole symmetric
and isotropic, which helps us to simplify calculations.

We consider a semi-infinite 3D TI material, such as Bi$_2$Se$_3$,
occupying the region $z>0$. The material boundary located at $z=0$
is perfectly flat and displays translational symmetry in the $(x,y)$
plane. The potential at the surface of a real 3D TI material is
different from the bulk crystal potential, irrespective of whether
the surface is kept in ultra-high vacuum or, for example, coated
with an overlayer or interfaced with another material. To
demonstrate the effect of the surface modification on the
topological states within a conceptually simple scheme, we introduce
the interaction of electrons with an external perturbation confined
at the surface, implementing the effective SP
$\mathbb{U}(\mathbf{r})$ into the EF calculation. Thus we write the
full electron energy of the truncated 3D TI in the following form:
\begin{equation}\label{Omega}
\Omega=\int_{z>0} d\mathbf{r} \Theta^{\dag}(\mathbf{r})[\mathbb{H}
(-i\nabla)+\mathbb{U}(\mathbf{r})]\Theta(\mathbf{r}),
\end{equation}
Here the operator $\mathbb{H}(-i\nabla)$ determined in Eq.
(\ref{H-TI}) acts in the the spinor function space
$\Theta(\mathbf{r})=(\theta_{1}(\mathbf{r}),
\theta_{2}(\mathbf{r}),\theta_{3}(\mathbf{r}),\theta_{4}
(\mathbf{r}))^{\mathrm{tr}}$,  represented in the basis
$u_{\mathbf{k}_{0}}$, the superscript $\mathrm{tr}$ denotes the
transpose operation. The EF components $\theta_{j}(\mathbf{r})$
(the subscript $j$ numbers the spinor components) are presumed to
be smooth and continuous functions in the half-space $z>0$, while
the spatial symmetry and periodicity of the system are broken due
to existence of the TI surface. It is evident that the
$\mathbf{k}\mathbf{p}$ approach cannot provide a correct
description of the wave-function behavior near the surface, where
large momenta are highly important. To overcome this drawback we
introduce the effective SP $\mathbb{U}(\mathbf{r})$, which affects
the electron states of TI at the surface. The potential
$\mathbb{U}(\mathbf{r})$ is nonzero in a small region $d$ (of the
order of a lattice parameter) around the geometrical boundary
$z=0$, where the validity of the $\mathbf{k}\mathbf{p}$ scheme is
questionable. An introduction of the phenomenological SP in
Eq.~(\ref{Omega}) enables us to correctly match the low-energy and
long-range electronic states inside the truncated TI with
evanescent vacuum states through the boundary conditions for EF
$\Theta(\mathbf{r})$. As long as the EF spatial variation of the
sought state,
$\Theta(\mathbf{r})=\sum_{\bm{\kappa}}\Theta(\bm{\kappa},z)\exp(i\bm{\kappa}\bm{\rho})$
[$\bm{\rho}=(x,y)$, $\bm{\kappa}=(k_{x},k_{y})$], is sufficiently
slow in the direction normal to the surface, one can adopt a local
approximation for the SP. Namely one writes
$\mathbb{U}(\mathbf{r})=d\mathbb{U}(\bm{\rho})\delta(z+0)$, where
the symbol $+0$ at the argument of the delta-function signifies
that the sheet-like SP is placed inside the TI half-space but at
infinitesimally small distance from the boundary $z=0$.

As a matter of course, an electron wave function has to be
continuous at a crystal boundary. Nevertheless, in the system
under consideration, since the Bloch factors of the wave function
inside and outside TI do not coincide (in particular, they have
distinct space symmetries), the long-range EF
$\Theta(\bm{\kappa},z)$ can formally undergo a finite break (jump)
across the boundary from $z=0-$ to $z=0+$ within the utilized
$\mathbf{kp}$ method (we refer the reader to the detailed
discussion in Ref.~\cite{Medhi}). In the current work, we do not
care how the wave-function behaves in the half-space $z<0$ but
next we make use of a functional
\begin{equation}\label{F}
F\{\Theta^{\dag},\Theta\}=\int_{0}^{\infty} dz
\Theta^{\dag}(\bm{\kappa},z)[\mathbb{H}
(\bm{\kappa},-i\partial_{z})+d
\mathbb{U}(\bm{\kappa})\delta(z+0)-\mathbb{I}E]\Theta(\bm{\kappa},z)
,
\end{equation}
where the energy $E$ plays a role of the Lagrange multiplier,
$\mathbb{I}$ is an unit $4\times4$ matrix,
$\partial_{z}=\partial/\partial z$. The functional (\ref{F}) is
determined in the class of the smooth and continuous EFs in the TI
half-space $z>0$ and includes the effective surface potential
$d\mathbb{U}(\bm{\kappa})\delta(z+0)$. Since, in a plane geometry,
the wave-vector $\bm{\kappa}$ is a good quantum number, we determine
the functional for each EF $\bm{\kappa}$-mode,
$\Theta(\bm{\kappa},z)$. Varying functional
$F\{\Theta^{\dag},\Theta\}$ with respect to $\Theta^{\dag}$ yields
the Euler equations for the half-space $z>0$ and the boundary
conditions at the surface at $z=0+$. The corresponding equations in
the compact form are:
\begin{equation}\label{Equation}
[\mathbb{H}
(\bm{\kappa},-i\partial_{z})-\mathbb{I}E]\Theta(\bm{\kappa},z)=0,
\end{equation}
\begin{equation}\label{Current}
i\frac{\delta
\mathbb{H}(\bm{\kappa},-i\partial_{z})}{\delta(-i\partial_{z})}\Theta(\bm{\kappa},z)|_{z=0+}
=2d\mathbb{U}(\bm{\kappa})\Theta(\bm{\kappa},z)|_{z=0+}.
\end{equation}
In the left side of Eq.~(\ref{Current}) the current density operator
acts on the EF spinor. Thus the right side associated with the
surface perturbation plays a role of the external (with regard to
the TI bulk) current source (sink). The equation (\ref{Current})
involves the surface potential parameters, in this sense it has
something in common with the equation which was used to calculate
the surface states of a crystal with a relativistic band structure
in Ref.~\cite{Volkov}. The solution of the boundary task,
Eqs.~(\ref{Equation}) and (\ref{Current}), answers the principal
physical question how the perturbation located just at the TI
boundary affects the near-surface topological states.

In the half-space $z>0$, the general solution of
Eq.~(\ref{Equation}) for each EF spinor component obeying the
condition $\theta_{j}(\bm{\kappa},z\rightarrow\infty)=0$ can be
represented as
\begin{equation}\label{EF}
\theta_{j}(\bm{\kappa},z)=\theta_{j}^{0}(\phi)\{\alpha_{j}(\kappa,E)\exp[-q_{1}(\kappa,E)z]
+\beta_{j}(\kappa,E)\exp[-q_{2}(\kappa,E)z]\},
\end{equation}
where
\begin{equation}\label{q}
q_{1,2}(\kappa,E)=\sqrt{q_{1,2}^{2}(E)+\kappa^{2}},
\end{equation}
\begin{equation}\label{q1,2}
q_{1,2}^{2}(E)=\frac{\mathrm{A}^{2}-2\mathrm{B}\Xi\pm\sqrt{\mathrm{A}^{4}-
4\mathrm{B}\Xi\mathrm{A}^{2}+4\mathrm{B}^{2}E^{2}}}{2\mathrm{B}^{2}}.
\end{equation}
Here the phase factors $\theta_{j}^{0}(\phi)$ forming the spinor
$\Theta^{0}(\phi)= (i, -sgn(\mathrm{A}), \mp e^{i\phi}, \pm
sgn(\mathrm{A})i e^{i\phi})^{\mathrm{tr}}$ depend only on the
momentum polar angle, $\phi$, $k_{x}\pm i k_{y}=\kappa\exp(\pm
i\phi)$; the signs $\pm$ relates to lower and upper spectral
branches, respectively. The characteristic momenta
$q_{1,2}(\kappa,E)$ are the solutions of the  corresponding secular
equation; $\kappa=|\bm{\kappa}|$. The boundary conditions,
Eq.~(\ref{Current}), determine the coefficients
$\alpha_{j}(\bm{\kappa},E)$ and $\beta_{j}(\bm{\kappa},E)$ as well
as the dispersion relation for the near-surface states inside the
bulk band gap, $|E(\kappa)|<\Xi$. The parameter
$\lambda=\mathrm{A}^{2}/4\mathrm{B}\Xi$ is implied to be
$\lambda\geqslant1$.

\section{Near-surface topological bound states}

The potential $\mathbb{U}$ in Eq.~(\ref{Current}) is a $4\times4$
matrix specifying internal properties of the TI surface and the
matrix elements include different components of scattering of the TI
states on SP. In principle, choosing the structure of the matrix and
the strength of its components allows us to tune spatial and energy
characteristics of the topological states. For example, as for the
SP diagonal matrix elements, $U_{jj}$, the values
$U_{1}=(U_{11}+U_{33})/2$ and $U_{2}=(U_{22}+U_{44})/2$ are
proportional to the scattering intensity of particle and hole,
respectively, on the spin-independent part of SP, while the
quantities $Q_{1}=(U_{11}-U_{33})/2$ and $Q_{2}=(U_{22}-U_{44})/2$
are proportional to the scattering intensity of particle and hole,
respectively, on the $z$-component of the exchange part of SP. The
off-diagonal matrix elements $U_{jj'}$ with $j\neq j'$ result from
the spin-orbit interaction at the surface, which, in general
different from the SOC in the TI bulk.

In this work we focus on the SP that preserves time reversal
symmetry, i.e. $\mathbb{U}=diag\{U_{1},U_{2},U_{1},U_{2}\}$, where
$U_{11}=U_{33}=U_{1}$, $U_{22}=U_{44}=U_{2}$. Such the SP structure
in the basis $u_{\mathbf{k}_{0}}$ results in the following relations
between the EF coefficients in Eq.~(\ref{EF}):
$\alpha_{3}=\alpha_{1}$, $\beta_{3}=\beta_{1}$,
$\alpha_{4}=\alpha_{2}$, $\beta_{4}=\beta_{2}$. Moreover, we neglect
the dependence of $U_{1,2}(\bm{\kappa})$ on $\bm{\kappa}$ in
Eq.~(\ref{Current}).

One can interpret the spin-independent scattering on the surface
within the framework of a 'local band bending' scheme (which is
quite reasonable for the contact of two insulators/semiconductors),
where the band edge corresponding to the $j$-th spinor component is
affected by the external perturbation confined at the surface:
$\Xi\rightarrow\Xi\pm dU_{j}\delta(z)$. In the situation of TI
covered with an overlayer the intuitive idea is that the diagonal
components of SP could be heuristically adjusted to the relative
offsets between the corresponding energy levels (bands) of the TI
substrate and the overlayer. In other words, the energy
$dU_{j}\delta(z)$ mimics the local bending of the respective bands.

After some algebra the corresponding secular equation results in the
implicit relation between the energy $E$ and the in-plane momentum
$\kappa$ for the bound state at the TI surface:
\begin{eqnarray}\label{Evskappa}
&&\mathrm{A}^{2}[q_{1}\pm\kappa][q_{2}\pm\kappa]
-[\mathrm{B}q_{1}^{2}+\Xi(\kappa)-E][\mathrm{B}q_{2}^{2}+\Xi(\kappa)-E]\nonumber\\&+&
2\biggl\{q_{1}q_{2}-\frac{\Xi(\kappa)-E}{\mathrm{B}}\pm \kappa
[q_{1}+q_{2}]\biggr\} \times
\biggl\{\mathrm{B}^{2}q_{1}q_{2}-dU_{1}dU_{2}-\frac{\mathrm{A}^{2}}{4}\biggr\}\\&-&
2[\Xi(\kappa)-E][dU_{2}(q_{1}+q_{2})\pm\kappa(dU_{1}+dU_{2})]\nonumber\\&-&
2\mathrm{B}\{dU_{1}q_{1}q_{2}(q_{1}+q_{2})\pm \kappa
dU_{1}(q_{1}+q_{2})^{2}\mp
\kappa(dU_{1}+dU_{2})q_{1}q_{2}\}=0\nonumber,
\end{eqnarray}
where $q_{1,2}=q_{1,2}(\kappa,E)$ in accordance with
Eqs.~(\ref{q}) and (\ref{q1,2}),
$\Xi(\kappa)=\Xi-\mathrm{B}\kappa^{2}$. Note that
Eq.~(\ref{Evskappa}) is invariant under the simultaneous
permutations: $E\leftrightarrow -E$,
$\kappa\leftrightarrow-\kappa$ and $U_{1}\leftrightarrow-U_{2}$.

At the $\Gamma$ point, Eq.~(\ref{Evskappa}) is reduced to the
equation that determines the Dirac (node) point position,
$E(\kappa=0)=E_{0}(U_{1,2})$, as a function of the SP strength
$U_{1,2}$:
\begin{eqnarray}\label{EvsU}
&&\biggl[\sqrt{1+\frac{E}{\Xi}}-\sqrt{1-\frac{E}{\Xi}}\biggr]
\biggl[\lambda+\sqrt{1-\frac{E^{2}}{\Xi^{2}}}-\frac{dU_{1}dU_{2}}{\mathrm{B}\Xi}\biggr]\nonumber\\&-&
\sqrt{\frac{2}{\mathrm{B}\Xi}}\biggl[dU_{1}\sqrt{1+\frac{E}{\Xi}}+dU_{2}\sqrt{1-\frac{E}{\Xi}}\biggr]
\times  \sqrt{2\lambda-1+\sqrt{1-\frac{E^{2}}{\Xi^{2}}}}=0.
\end{eqnarray}

\begin{figure}
\includegraphics[width=\columnwidth]{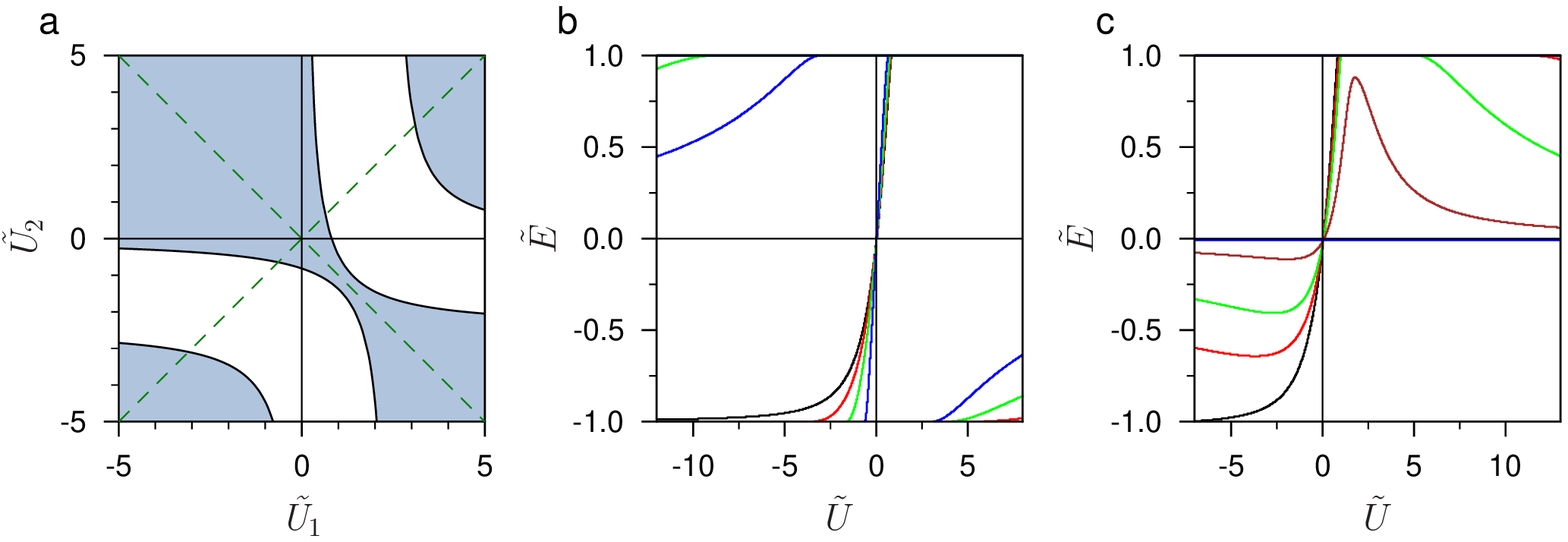}
\caption{(Color online) (a) Realm of the near-surface state Dirac
point position in a 3D TI in terms of the matrix elements $U_{1}$
and $U_{2}$. In the dashed areas the Dirac point is inside the bulk
band gap. (b) Position of the Dirac point versus the SP strength
when $U=U_{1}=mU_{2}$, where $m=0$ (black line), $m=0.1$ (red line),
$m=0.3$ (green line),  $m=1.0$ (blue line). (c) Position of the
Dirac point versus the SP strength when $U=U_{1}=mU_{2}$, where
$m=0$ (black line), $m=-0.2$ (red line), $m=-0.4$ (green line),
$m=-0.8$ (brown line), $m=-1.0$ (blue line coinciding with an
abscissa axis). The units used are,
$\tilde{U}=\frac{dU}{\sqrt{\mathrm{B}\Xi}}$,
$\tilde{E}=\frac{E}{\Xi}$, and
$\lambda=\frac{\mathrm{A}^{2}}{4\mathrm{B}\Xi}=2$.}
 \label{fig1}
\end{figure}

The shaded areas in Fig.~\ref{fig1}a denote the realm of the
near-surface bound state with $|E_{0}(U_{1,2})|<\Xi$ on the
$(U_{1},U_{2})$-plane. The dependence of the node point position
on the SP strength, obtained from Eq.~(\ref{EvsU}) for several
ratio values $U_{1}/U_{2}$, is plotted in Fig.~\ref{fig1}b (for
$(U_{1}U_{2})>0$) and Fig.~\ref{fig1}c (for $(U_{1}U_{2})<0$). One
can see three different regions in these plots. At weak potential
$d|U_{1,2}|<<\sqrt{\mathrm{B}\Xi}$, the Dirac point linearly
shifts with respect to the TI bulk bands to either higher or lower
binding energies depending on the SP strength sum,
\begin{equation}\label{E0U-small}
E_{0}(U_{1},U_{2})=\frac{2\sqrt{\lambda}}{1+\lambda}\sqrt{\frac{\Xi}{\mathrm{B}}}(dU_{1}+dU_{2}).
\end{equation}
In case the SP strength is large,
$d|U_{1}|+d|U_{2}|>>\sqrt{\mathrm{B}\Xi}$, the node point energy
approaches zero as
\begin{equation}\label{E0U-large}
E_{0}(U_{1},U_{2})=-2\sqrt{\lambda\mathrm{B}\Xi^{3}}\biggl(\frac{1}{dU_{1}}+\frac{1}{dU_{2}}\biggr).
\end{equation}
On the $(U_{1},U_{2})$-plane, there are regions (unshaded areas in
Fig.~\ref{fig1}a) where the bound state is absent since the node
point merges into the conduction or valence bulk band. For example,
if $U_{1}=U_{2}=U$, the threshold values of the potential, at which
the node point splits off the bulk band continuum, are
$2dU_{\pm}=\sqrt{\mathrm{B}\Xi}[\sqrt{2(4\lambda-1)}\pm\sqrt{2(2\lambda-1)}]$,
so that $E_{0}(\pm U_{-})=\pm\Xi$ and $E_{0}(\pm U_{+})=\mp\Xi$.

If the energy $\varepsilon(\kappa)$ is a small deviation from the
Dirac linear spectrum,
$E^{(\pm)}(\kappa)=\pm\mathrm{A}\kappa+\varepsilon(\kappa)$,
$|\varepsilon(\kappa)|<<\Xi$, the characteristic momenta,
Eq.~(\ref{q1,2}), are found as
\begin{equation}\label{q1,2plus}
q_{1,2}^{(\pm)}(\kappa,E)=q_{1,2}^{0}(\kappa)+
\frac{(\pm\kappa)\varepsilon(\kappa)}{\mathrm{B}q_{1,2}^{0}(\kappa)[q_{1,2}^{0}(\kappa)-q_{2,1}^{0}(\kappa)]},
\end{equation}
where $q_{1,2}^{0}(\kappa)$ is given by Eq.~(\ref{Q0-12}). The
deviation $\varepsilon(\kappa)$ appears to be small not only when
$U_{1}\simeq-U_{2}$ but also when SP is either weak or strong. Using
the expression (\ref{q1,2plus}) one can obtain the spectrum and
estimate the spatial distribution of the near-surface state in these
limit situations.

So, for the extremely large potential,
$|U_{1,2}|\rightarrow\infty$, the correction approaches zero,
$\varepsilon(\kappa)\rightarrow0$, in turn, the EF coordinate
dependence is described by a difference of the exponents,
$\Theta(\bm{\kappa},z)\sim\exp[-q_{1}^{0}(\kappa)z]-\exp[-q_{2}^{0}(\kappa)z]$,
so that the maximum of the electron density, $|\Theta(z)|^{2}$,
does not occur on the surface, where $\Theta(z=0)=0$, but rather
near the point
$z_{0}=\ln(q_{1}^{0}/q_{2}^{0})/(q_{1}^{0}-q_{2}^{0})$ (where
$z_{0}\lesssim \sqrt{\frac{\mathrm{B}}{\Xi}}<(q_{2}^{0})^{-1}$)
that is distant from the surface. Such the EF distribution,
together with the linear spectrum, was found under the free
boundary conditions~\cite{Shan}. Our approach allows us to capture
peculiarities of the surface state in 3D TI induced by the SP. If
the SP strength is much greater than the characteristic energy,
$d|U_{1,2}|\gg\mathrm{B}q_{1}^{0}$, within the perturbation
theory, one obtains the amendment to the linear dispersion law as
\begin{equation}\label{corr-U-large}
\varepsilon(\kappa)=-\frac{|\mathrm{A}|\Xi^{2}}{\Xi(\kappa)}
\biggl(\frac{1}{dU_{1}}+\frac{1}{dU_{2}}\biggr).
\end{equation}
The surface state spectrum acquires a curvature and a shift of the
node point, $E_{0}=\varepsilon(0)$, which are inversely
proportional to the potential, however the fermion group velocity
$|\mathrm{A}|$ near the node point does not change since the
amendment $\varepsilon(\kappa)$ (\ref{corr-U-large}) does not
contain a contribution linear in $\kappa$. In the lowest order in
$(U_{1,2})^{-1}$, the relations between the coefficients in
Eq.~(\ref{EF}) are given by
\begin{equation}\label{a-to-b-ratios}
\frac{\beta_{1}}{\alpha_{1}}=-1+\frac{\sqrt{\mathrm{A}^{2}-4\mathrm{B}\Xi(\kappa)}}{dU_{1}},~
\frac{\beta_{2}}{\alpha_{2}}=-1-\frac{\sqrt{\mathrm{A}^{2}-4\mathrm{B}\Xi(\kappa)}}{dU_{2}},
\end{equation}
Thus, the electron density does not vanish on the TI surface,
$|\Theta(z=0)|^{2}\sim(U_{1,2})^{-2}$. However, under the SP
influence, the EF components can vanish near the surface at
$z=z_{1,2}<z_{0}$, namely, $\theta_{1,3}(0,z)=0$ at
$z=z_{1}=\mathrm{B}/dU_{1}$ when $U_{1}>0$, and
$\theta_{2,4}(0,z)=0$ at $z=z_{2}=-\mathrm{B}/dU_{2}$ when
$U_{2}<0$. Besides, as seen from Eq.~(\ref{q1,2plus}), the SP
affects the decay length of the EF nonzero harmonics.

If the SP is formally absent, $U_{1,2}=0$, one arrives at the solution
obtained in Ref.~\cite{Medhi} from using the natural boundary
conditions: the surface state shows the linear spectrum
$E^{(\pm)}(\kappa)=\pm\mathrm{A}\kappa$ and the EF spatial profile
in $z$-direction is merely a sum of the two exponents,
$\Theta(\bm{\kappa},z)\sim\exp[-q_{1}^{0}(\kappa)z]+\exp[-q_{2}^{0}(\kappa)z]$,
i.e., the probability density of the near-surface state is peaked on
the boundary $z=0$ and its tail penetrates into the TI bulk with the
decay length $(q_{2}^{0})^{-1}$. In the case of weak SP,
$d|U_{1,2}|\ll\sqrt{\mathrm{B}\Xi}$, the correction to the
dispersion law is given by
\begin{equation}\label{corr-U-small}
\varepsilon(\kappa)=
\frac{4|\mathrm{A}|\Xi(\kappa)}{\mathrm{A}^{2}+4\mathrm{B}\Xi(\kappa)}(dU_{1}+dU_{2}).
\end{equation}
The spin-independent SP is seen to entirely shift and warp the
energy-momentum dependence. Note, that the corrections
(\ref{corr-U-large}) and (\ref{corr-U-small}) are opposite in the
sign. Turning on the SP leads to the different contributions of the
quick and slow exponents into EF:
$\beta_{j}/\alpha_{j}=1+o(U_{1,2})$.

\begin{figure}
\includegraphics[width=\columnwidth]{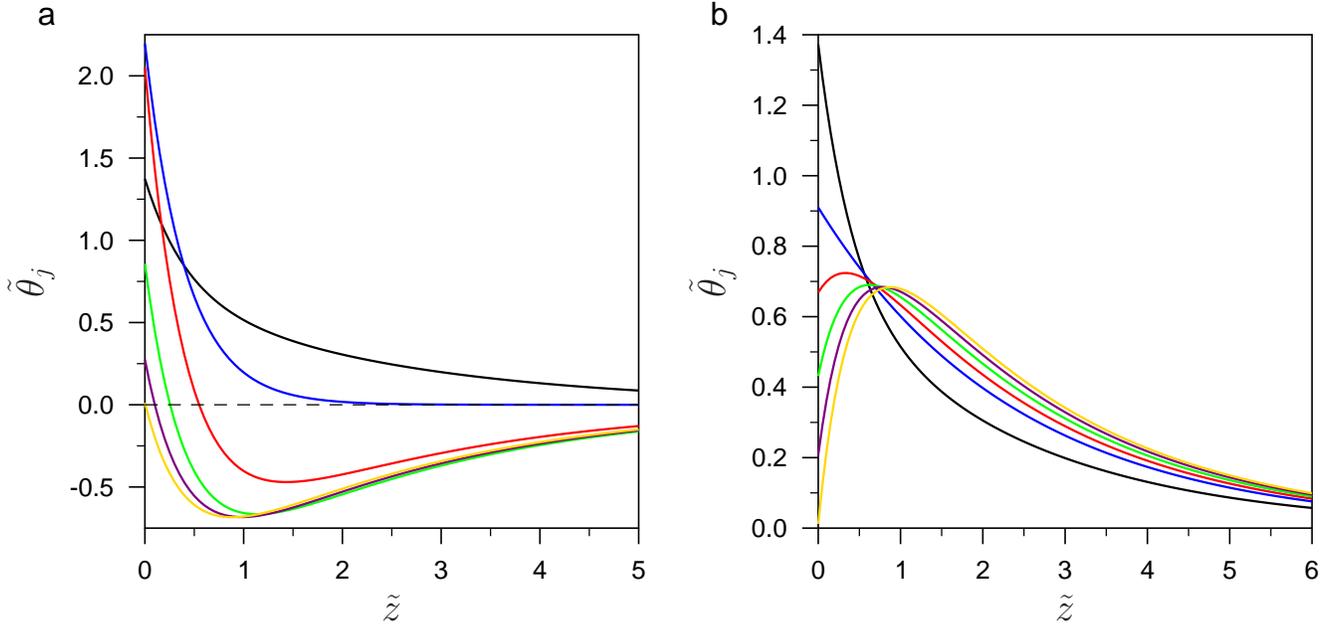}
\caption{(Color online)  Space dependence of the zeroth harmonic
($\kappa=0$) of the envelope function $\theta_{j}(\kappa,z)$
Eq.~(\ref{EF-stagg}) in the case of $U_{1}=-U_{2}=U$ for several
values of the potential: (a) $\tilde{U}=0$ (black line),
$\tilde{U}=1$ (blue line), $\tilde{U}=2$ (red line), $\tilde{U}=4$
(green line), $\tilde{U}=10$ (brown line), $\tilde{U}=\infty$
(yellow line); (b) $\tilde{U}=0$ (black line), $\tilde{U}=-1$ (blue
line), $\tilde{U}=-2$ (red line), $\tilde{U}=-4$ (green line),
$\tilde{U}=-10$ (brown line), $\tilde{U}=-\infty$ (yellow line);
where $\tilde{U}=\frac{dP}{\sqrt{\mathrm{B}\Xi}}$,
$\tilde{z}=\sqrt{\frac{\Xi}{\mathrm{B}}}z$,
$\tilde{\theta}_{j}=\sqrt[4]{\frac{\mathrm{B}}{\Xi}}\theta_{j}$,
$\lambda=\frac{\mathrm{A}^{2}}{4\mathrm{B}\Xi}=2$. The envelope
function is normalized as
$\int_{0}^{\infty}dz|\theta_{j}(\bm{\kappa},z)|^{2}=1$.}
 \label{fig2}
\end{figure}

Let us consider thoroughly the specific case of the staggered
alignment of the matrix elements, $U_{1}=-U_{2}=U$, when SP does
not break the particle-hole symmetry. One can verify in
Eq.~(\ref{Evskappa}) that in such the case the near-surface state
maintains the ideal Dirac spectrum
$E^{(\pm)}(\kappa)=\pm\mathrm{A}\kappa$ regardless of the size and
sign of $U$. While the spectrum is independent of the SP, the
envelope function is strongly affected by it. The coordinate
dependence of each component of the EF spinor is given by
\begin{equation}\label{EF-stagg}
\theta_{j}(\kappa,z)=\theta_{j}^{0}\sqrt{\frac{\Xi}{\mathrm{B}}}
\frac{(1+\nu)\exp(-q_{1}^{0}z)+(1-\nu)\exp(-q_{2}^{0}z)}
{\sqrt{\frac{(1+\nu)^{2}}{2}q_{2}^{0}+\frac{(1-\nu)^{2}}{2}q_{1}^{0}+
(1-\nu^{2})\frac{2\Xi(\kappa)}{|\mathrm{A}|}}},
\end{equation}
where
\begin{equation}\label{Q0-12}
q_{1,2}^{0}=q_{1,2}^{0}(\kappa)=q_{1,2}(\kappa,|E|=|\mathrm{A}|\kappa)=
\frac{|\mathrm{A}|\pm\sqrt{\mathrm{A}^{2}-4\mathrm{B}\Xi(\kappa)}}{2\mathrm{B}},
\end{equation}
\begin{equation}\label{nu}
\nu=\nu(\kappa)=\frac{2dU}{\sqrt{\mathrm{A}^{2}-4\mathrm{B}\Xi(\kappa)}},
\end{equation}
The EF of Eq.~(\ref{EF-stagg}) is normalized as
$\int_{0}^{\infty}|\theta_{j}(\kappa,z)|^{2}=1$. The spatial
behavior of the EF zeroth harmonic $\theta_{j}(0,z)$ is
illustrated in Fig.~\ref{fig2}a and Fig.~\ref{fig2}b for positive
and negative $U$, respectively. With increasing SP strength the EF
structure evolves from the sum of the exponents at $U=0$ (black
lines) to the difference at $|U|\rightarrow\infty$ (yellow lines).
So, one sees a gradual change in the profile of the near-surface
state such that its gravity center moves from the surface to the
TI interior. It is obvious the behavior of the EF nonzero
harmonics Eq.~(\ref{EF-stagg}) is insignificantly different from
what is plotted in Fig.~\ref{fig2}; unless the tail at $\kappa\neq
0$ is slightly longer than that at $\kappa=0$.

\begin{figure}
 \begin{center}
\includegraphics[width=0.5\columnwidth]{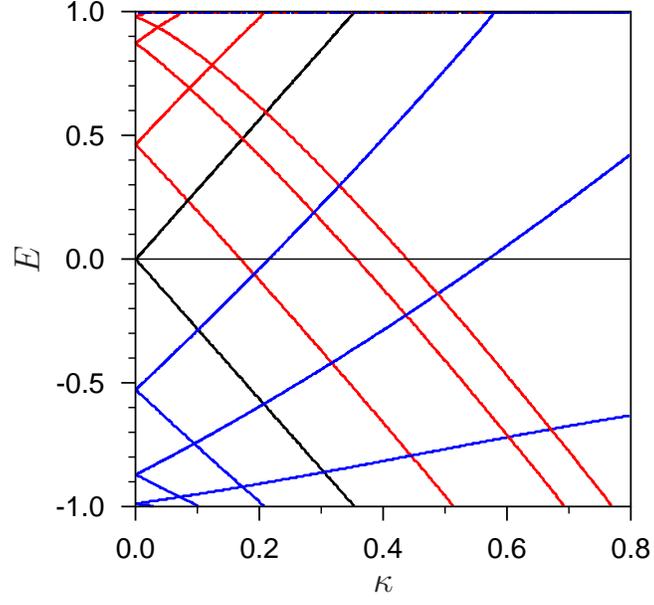}
\caption{(Color online) Spectrum of surface states $E(\kappa)$ at
$U_{1}=U_{2}=U$ for several values of SP:  $\tilde{U}=0$ (black
line), $\tilde{U}=0.25, 0.5, 0.6$ (red lines), $\tilde{U}=3.5, 5.0,
10.0$ (blue lines); $\tilde{U}=\frac{dU}{\sqrt{\mathrm{B}\Xi}}$,
$\tilde{E}=\frac{E}{\Xi}$,
$\tilde{\kappa}=\sqrt{\frac{\mathrm{B}}{\Xi}}\kappa$,
$\lambda=\frac{\mathrm{A}^{2}}{4\mathrm{B}\Xi}=2$.}
 \label{fig3}
 \end{center}
\end{figure}

To study the modification of the near-surface states under the
finite strength of SP we dwell at length on the situation
$U_{1}=U_{2}=U$. As seen in Fig.~\ref{fig3}, at a finite strength of
$U$, $d|U|\simeq\sqrt{\mathrm{B}\Xi}$, apart from the aforesaid
shift of the node point, the form of the spectral dependence,
$E(\kappa)$, alters (in comparison with the limiting cases $U=0$ or
$U\rightarrow\pm\infty$) under the SP influence. The group velocity
of the surface topological excitations decreases from the quantity
$|\mathrm{A}|$ to zero when the SP strength $|U|$ either increases
from zero to the threshold value $U_{-}$ or decreases from infinity
to the threshold value $U_{+}$. A noticeable deviation from
linearity can be seen for the strength
$d|U|\simeq\sqrt{\mathrm{B}\Xi}$. In the limit $|U|\rightarrow
U_{-}$ the dispersion becomes parabolic at small $\kappa$, and in
the limit $|U|\rightarrow U_{+}$ the curve $E(\kappa)$ smoothly
merges into $E=-\Xi$. The dependence $E(\kappa)$ acquires a
curvature so that the relatively strong ($|U|>U_{+}$) and relatively
weak ($|U|<U_{-}$) potentials provide with the curvature of opposite
sing.

\begin{figure}
\includegraphics[width=\columnwidth]{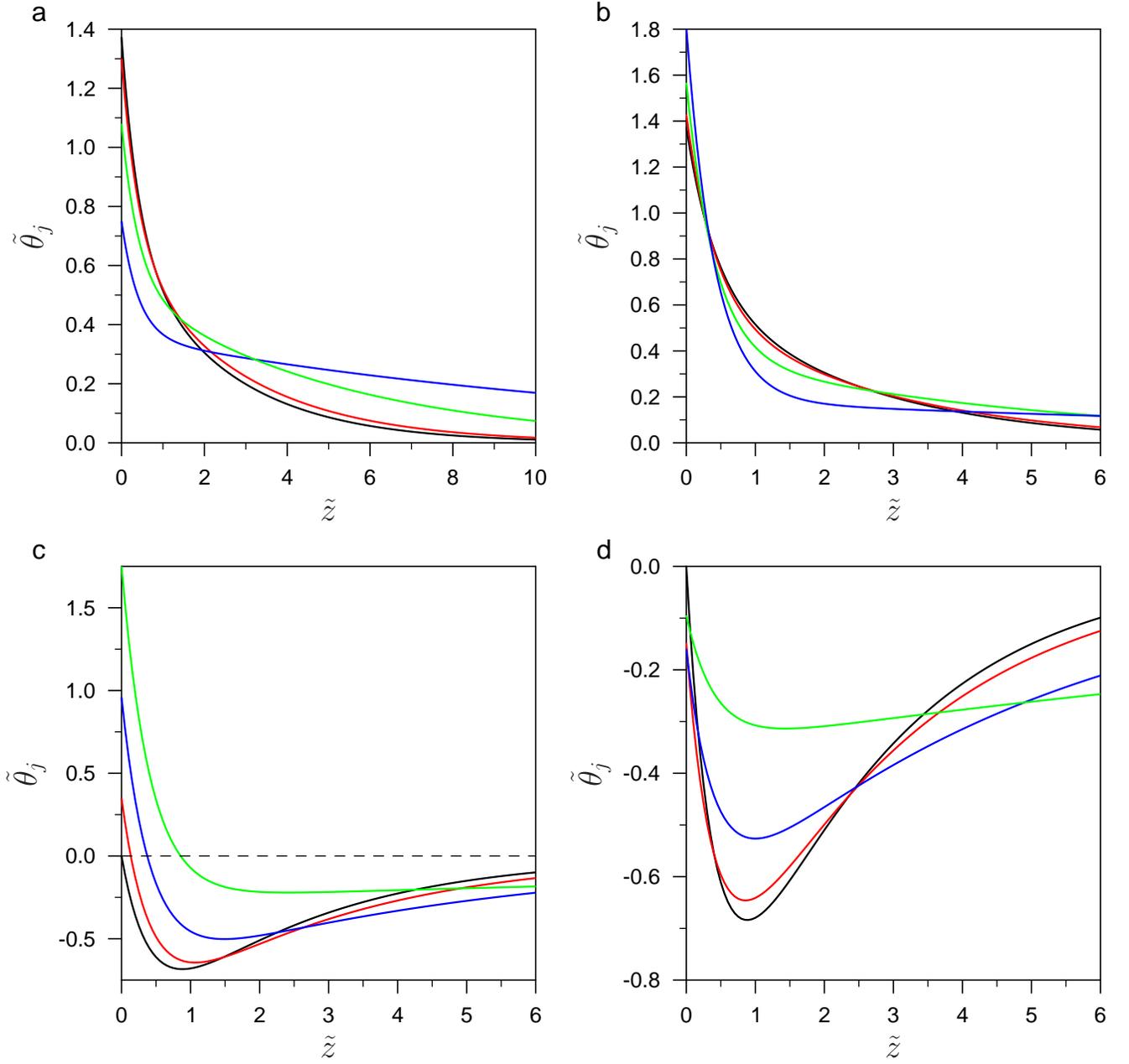}
\caption{(Color online)  Space dependence of the zeroth harmonic
($\kappa=0$) of the envelope function $\theta_{j}(\kappa,z)$ in the
case of $U_{1}=U_{2}=U$ for several values of (a) the relative weak
positive potential with $0<U<U_{-}$ [$\tilde{U}=0$ (black line),
$\tilde{U}=0.25$ (red line), $\tilde{U}=0.5$ (green line),
$\tilde{U}=0.6$ (blue line)], (b) the relative weak negative
potential with $-U_{-}<U<0$ [$\tilde{U}=0$ (black line),
$\tilde{U}=-0.25$ (red line), $\tilde{U}=-0.5$ (green line),
$\tilde{U}=-0.6$ (blue line)], (c) the relative large positive
potential with $U_{+}<U<\infty$ [$\tilde{U}=\infty$ (black line),
$\tilde{U}=10$ (red line), $\tilde{U}=5$ (blue line),
$\tilde{U}=3.5$ (green line)], (d) the relative large negative
potential with $-\infty<U<-U_{+}$ [$\tilde{U}=-\infty$ (black line),
$\tilde{U}=-10$ (red line), $\tilde{U}=-5$ (blue line),
$\tilde{U}=-3.5$ (green line)], where
$\tilde{U}=\frac{dU}{\sqrt{\mathrm{B}\Xi}}$,
$\tilde{z}=\sqrt{\frac{\Xi}{\mathrm{B}}}z$,
$\tilde{\theta}_{j}=\sqrt[4]{\frac{\mathrm{B}}{\Xi}}\theta_{j}$,
$\lambda=\frac{\mathrm{A}^{2}}{4\mathrm{B}\Xi}=2$. The envelope
function is normalized as
$\int_{0}^{\infty}dz|\theta_{j}(\bm{\kappa},z)|^{2}=1$.}
 \label{fig4}
\end{figure}

The spatial behavior of the near-surface states for $U_{1}=U_{2}=U$
is shown in Fig.~\ref{fig4}. When the SP is weak, $0<|U|<U_{-}$, the
probability density $\sim |\theta_{j}(z)|^{2}$ is largely peaked
near the surface. The strong SP, $|U|>U_{+}$, pushes the probability
density towards the TI bulk. The EF is exponentially decaying away
from the surface. We would like to emphasize that the EF decay
lengths, $(q_{1,2})^{-1}$ (see Eq.~(\ref{q1,2})), are strongly
influenced by the SP strength. For example, when the strength $U$
varies either from $0$ to $\pm U_{-}$ or from $\pm\infty$ to $\pm
U_{+}$, the momentum $q_{1}(E)$ increases from
$\sqrt{\frac{\Xi}{\mathrm{B}}}(\sqrt{\lambda}+\sqrt{\lambda-1})$ to
$\sqrt{\frac{\Xi}{\mathrm{B}}}\sqrt{2(2\lambda-1)}$ (i.e. the 'long'
exponent of EF (\ref{EF})) becomes longer) and the momentum
$q_{2}(E)$ decreases from
$\sqrt{\frac{\Xi}{\mathrm{B}}}(\sqrt{\lambda}-\sqrt{\lambda-1})$ to
$0$ (i.e. the 'short' exponent becomes shorter).

\section{Quasi-topological bound states near the surface of a normal insulator}

Next, we investigate the effect of the surface modification on the
near-surface bound states when the bulk is a 3D normal
(topologically trivial) insulator. Here we address the fundamental
question of whether 3D NI responds to a localized surface
perturbation in a way different from 3D TI. In order to describe 3D
NI, one uses the same relativistic Hamiltonian~(\ref{H-TI}) in
which, however, now there is the normal (non-inverted) alignment of
the energy terms of different parity $|+\uparrow(\downarrow)\rangle$
and $|-\uparrow(\downarrow)\rangle$ around $\mathbf{k}_{0}=0$ is
implied, i.e., $\Xi<0$ and $\Xi\mathrm{B}<0$. For instance, in the
case of the
%Sb$_2$Te$_3$
In$_2$Se$_3$ crystal, which shares the same crystal structure with
Bi$_2$Se$_3$, the SOC is not strong enough to provide the
inversion between two $p_{z}$ orbitals with opposite parity at the
$\Gamma$ point~\cite{Zhang}. In the limit $\mathrm{A}\rightarrow
0$, when SOC is negligible small, Eq.~(\ref{H-TI}) defines merely
the semiconductor with simple (nonrelativistic) two-band spectrum.

\begin{figure}
\includegraphics[width=\columnwidth]{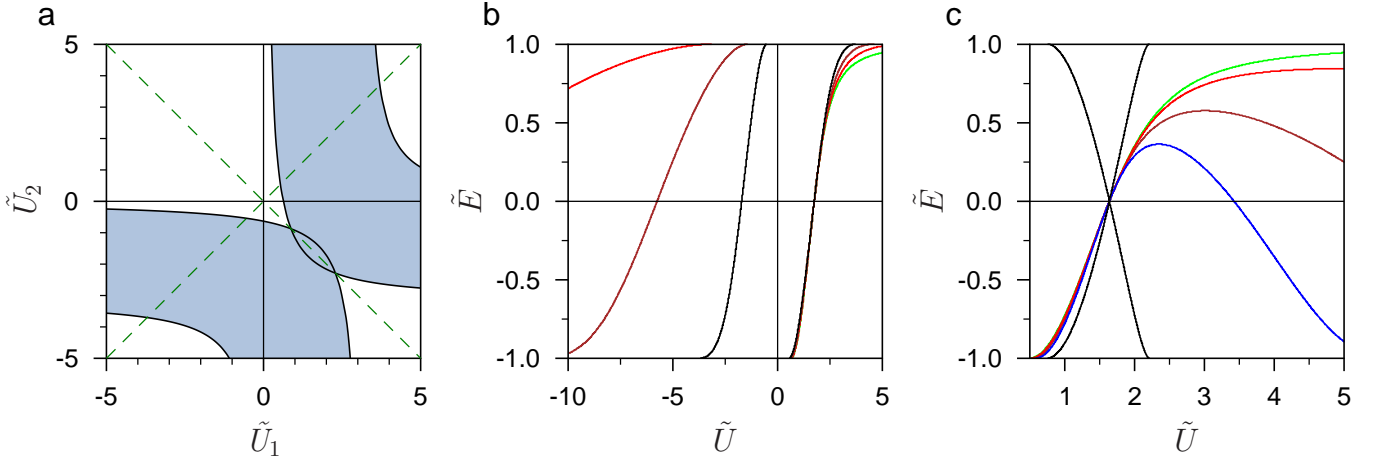}
\caption{(Color online) (a) Realm of the near-surface state
existence in 3D NI under SP with the matrix elements $U_{1}$ and
$U_{2}$. In the dashed (fill, painted) areas (domains) on the
$(U_{1},U_{2})$-plane, the point $E(0)$ lives inside the bulk band
gap. (b) Position of the band edge energy of the near-surface state
versus the SP strengths of the same sign under the stipulation
$U_{2}=nU_{1}$ and $U_{1}=U$, where $n=0$ (green line), $n=0.1$ (red
line), $n=0.3$ (brown line), $n=1.0$ (black line). (c) Position of
the band edge energy of the near-surface state versus the SP
strengths of opposite signs under the stipulation $U_{2}=nU_{1}$ and
$U_{1}=U$, where $n=0$ (green line), $n=-0.1$ (red line), $n=-0.3$
(brown line), $n=-0.5$ (blue line), $n=-1.0$ (black line). The
dimensional units are used,
$\tilde{U}=\frac{dU}{\sqrt{\mathrm{B}|\Xi|}}$,
$\tilde{E}=\frac{E}{|\Xi|}$, and
$|\lambda|=\frac{\mathrm{A}^{2}}{4\mathrm{B}|\Xi|}=2$.}
 \label{fig5}
\end{figure}

It is evident that Eqs.~(\ref{Omega})-(\ref{Current}) and the
relevant sentences are valid regardless of the $\Xi$ sign. Therefore
analogously to what has been done in the previous Sections (the
details are omitted) one can obtain the characteristics of the bound
electron states on the NI surface subjected to the external
spin-independent influence. The existence of these states is
determined by real solutions of the corresponding secular equation
within the bulk gap, $E(0)$, which is given by Eq.~(\ref{Evskappa})
at $\kappa=0$ and $\Xi=-|\Xi|$. Fig.~\ref{fig5}a shows the existence
realm of the near-surface bound state, i.e., the area on the
$(U_{1},U_{2})$-plane where $|E_{0}|<|\Xi|$. The energy $E(0)$ as a
function of the SP strength for several ratio values $U_{1}/U_{2}$
is represented in Fig.~\ref{fig5}b (for $U_{1}U_{2}>0$) and
Fig.~\ref{fig5}c (for $U_{1}U_{2}<0$). As is seen, except for the
quadrant $(U_{1}<0,U_{2}>0)$, one may choose the ratio $U_{1}/U_{2}$
to match the SP which induces the bound electron state on the NI
surface.

\begin{figure}
\includegraphics[width=\columnwidth]{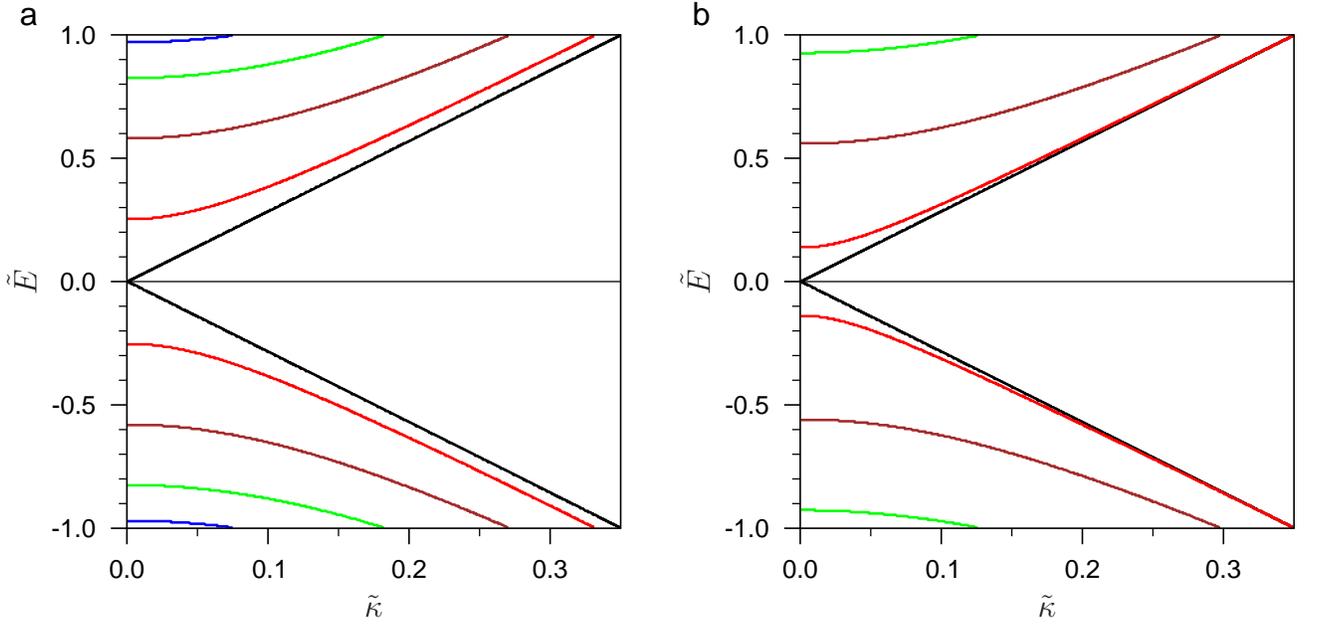}
\caption{(Color online)  Spectrum of the near-surface state
$E(\kappa)$ at $U_{1}=-U_{2}=U$ for several values of the SP
strength: (a) $U\leq U_{0}$ [$\tilde{U}=\sqrt{3}$ (black line),
$\tilde{U}=1.6$ (red line), $\tilde{U}=1.4$ (brown line),
$\tilde{U}=1.2$ (green line), $\tilde{U}=1.0$ (blue line)]; (b)
$U\geq U_{0}$ [$\tilde{U}=1.8$ (red line), $\tilde{U}=2.0$ (brown
line), $\tilde{U}=2.2$ (green line), $\tilde{U}=\sqrt{3}$ (black
line)], where $\tilde{U}=\frac{dU}{\sqrt{\mathrm{B}|\Xi|}}$,
$\tilde{E}=\frac{E}{|\Xi|}$,
$\tilde{\kappa}=\sqrt{\frac{\mathrm{B}}{|\Xi|}}\kappa$,
$|\lambda|=\frac{\mathrm{A}^{2}}{4\mathrm{B}|\Xi|}=2$.}
 \label{fig6}
\end{figure}

In what follows, we consider thoroughly only the two particular
cases: $U_{1}=-U_{2}$ and $U_{1}=U_{2}$. When the SP matrix elements
are in staggered rows, $U_{1}=-U_{2}=U$, the bound state exists at
$U>0$, which diminishes virtually the 3D NI bulk gap on the surface.
Because of the presence of such the SP, the particle-hole symmetry
of the system is preserved. The relations between the energy and
momentum is given by
\begin{equation}\label{NI-spectrum-1}
E(\kappa)= \pm\biggl\{
\Omega^{2}(\kappa)+\mathrm{A}^{2}\kappa^{2}- \biggl[dU\sqrt{
\frac{\mathrm{A}^{2}}{2\mathrm{B}^{2}}+\frac{2\Omega(\kappa)}{\mathrm{B}}
-\frac{d^{2}U^{2}}{\mathrm{B}^{2}}}-\frac{\mathrm{A}^{2}}{4\mathrm{B}}\biggr]^{2}
\biggr\}^{1/2} ,
\end{equation}
where $\Omega(\kappa)=|\Xi|+\mathrm{B}\kappa^{2}$,
$\pm\Omega(\kappa)$ is the projection of the bulk spectrum onto the
surface. In Fig.~\ref{fig6}, the spectral dependence $E(\kappa)$ is
illustrated for several choices of the SP strength $U$. So, the
system exhibits a non-linearly dispersing surface state which is
specified by the energy gap $2E(0)$. The crossing black lines in
Fig.~\ref{fig5}c show the half gap $E(0)$ as a function of $U$. One
can see in Fig.~\ref{fig5}a and Fig.~\ref{fig5}c that the
near-surface state stays in the bulk band gap, $|E(\kappa)|<|\Xi|$,
when the SP strength is restricted by the interval $U_{+}>U>U_{-}$,
where $\tilde{U}_{\pm}=(\sqrt{1+2|\lambda|}\pm 1)/\sqrt{2}$. While,
outside this interval, it is buried in the bulk band continuum. In
the close vicinity of a band-crossing point $\tilde{U}_{0}$, where
$|E(\kappa)|\ll |\Xi|$, the dispersion relation is given by
$E(\kappa)=\pm\sqrt{4\Xi^{2}(\tilde{U}-\tilde{U}_{0})^{2}+\mathrm{A}^{2}\kappa^{2}}$,
so that a degeneracy at the band-crossing point $U_{0}$ is lifted
due to gapping $\sim|U-U_{0}|$. It is convenient to measure the SP
strength in the dimensionless units
$\tilde{U}=\frac{dU}{\sqrt{\mathrm{B}|\Xi|}}$, then
$\tilde{U}_{0}=\sqrt{1+|\lambda|}$, where
$|\lambda|=\frac{\mathrm{A}^{2}}{4\mathrm{B}|\Xi|}$. In turn, in a
small energy window near the bulk band edges, where
$|\Xi|-|E(\kappa)|\ll |\Xi|$, the dependence (\ref{NI-spectrum-1})
becomes
\begin{equation}\label{NI-spectrum-2}
E(\kappa)= \pm\biggl\{
\Omega^{2}(\kappa)+\mathrm{A}^{2}\kappa^{2}-\frac{1}{4\lambda^{2}}
\biggl[|\Xi|(\tilde{U}_{+}^{2}-\tilde{U}^{2})(\tilde{U}^{2}-\tilde{U}_{-}^{2})
+2\tilde{U}^{2}\mathrm{B}\kappa^{2}\biggr]^{2} \biggr\}^{1/2}
.
\end{equation}
Note, in the case $\mathrm{A}=0$, Eq.~(\ref{NI-spectrum-1})
reduces to $E(\kappa)=\pm[\Omega(\kappa)-d^{2}U^{2}/\mathrm{B}]$;
in other words, in the 3D NI under finite value of SOC, a finite
strength of SP, $U_{-}\sim \mathrm{A}^{2}$, is required to split
off the in-gap near-surface state from the 3D bulk continuum.

Such the behavior of the bound state location in energy axis with
increasing the SP strength can intuitively be explained in the
language of the 'local band bending' scheme proposed in the previous
section. When 3D NI is brought into contact with a thin dielectric
overlayer, a relative weak positive SP, $U\gtrsim U_{-}$, splits off
states from both the conduction bulk band and valence one due to a
local narrowing of the gap, $|\Xi|-dU\delta(z)$. At $U\simeq U_{0}$,
the local band bending is so steep that the bulk band edges of 3D NI
cross over the energy levels of an adjusted overlayer with opposite
parity, as it would be if the 3D NI surface was in the contact with
a TI overlayer. Further increase of the strength $U$ above the
critical value $U_{+}$ pushes the near-surface state into the bulk
continuum.

\begin{figure}
\includegraphics[width=\columnwidth]{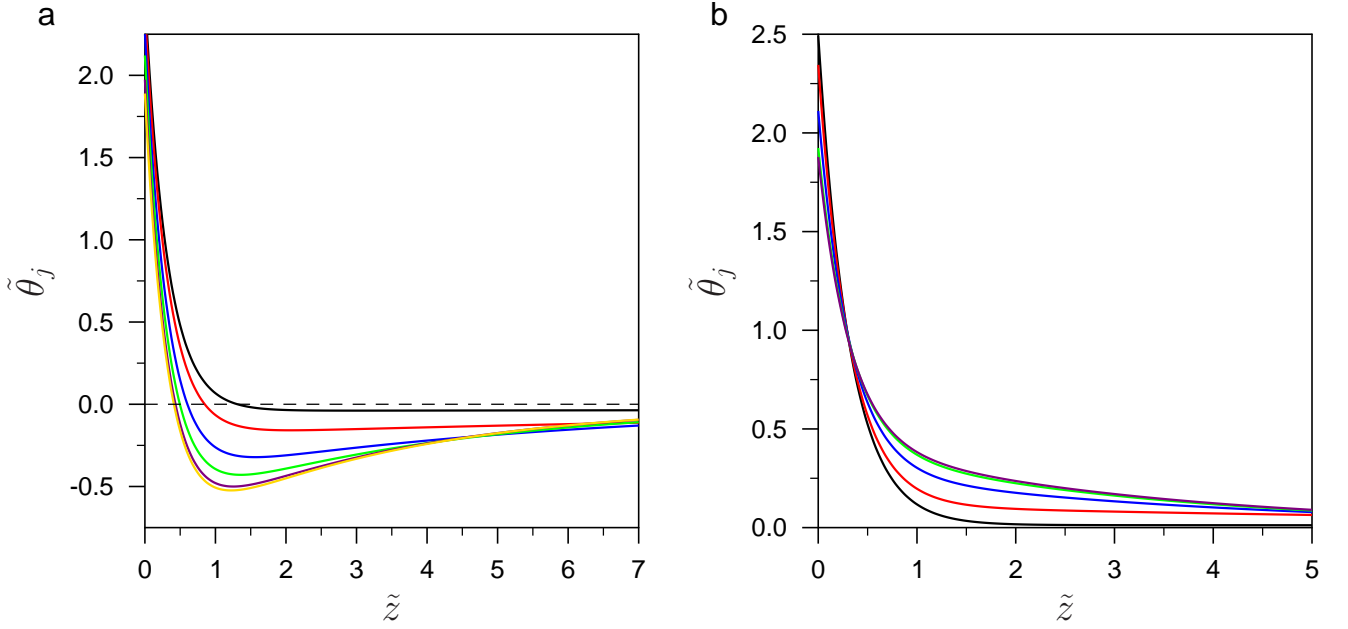}
\caption{(Color online)  Space dependence of the zeroth harmonic
($\kappa=0$) of the NI envelope function $\theta_{j}(\kappa,z)$ in
the case of $U_{1}=-U_{2}=U$ for several values of the SP strength
with (a) $U_{-}<U<U_{0}$ [$\tilde{U}=0.9$ (black line),
$\tilde{U}=1.0$ (red line), $\tilde{U}=1.2$ (blue line),
$\tilde{U}=1.4$ (green line), $\tilde{U}=1.6$ (brown line),
$\tilde{U}=1.7$ (yellow line)], (b) $U_{+}>U>U_{0}$
[$\tilde{U}=2.28$ (black line), $\tilde{U}=2.2$ (red line),
$\tilde{U}=2.0$ (blue line), $\tilde{U}=1.8$ (green line),
$\tilde{U}=1.75$ (brown line)], where
$\tilde{U}=\frac{dU}{\sqrt{\mathrm{B}|\Xi|}}$,
$\tilde{z}=\sqrt{\frac{|\Xi|}{\mathrm{B}}}z$,
$\tilde{\theta}_{j}=\sqrt[4]{\frac{\mathrm{B}}{|\Xi|}}\theta_{j}$,
$|\lambda|=\frac{\mathrm{A}^{2}}{4\mathrm{B}|\Xi|}=2$. The envelope
function is normalized as
$\int_{0}^{\infty}dz|\theta_{j}(\bm{\kappa},z)|^{2}=1$.}
 \label{fig7}
\end{figure}

Figure~\ref{fig7} visualizes the effect of the external potential
with $U_{1}=-U_{2}=U$, on the space profile of the in-gap state of
the truncated 3D NI. It is of interest to note that, while crossing
the value $U_{0}$, the form of the space dependence of the EF zeroth
harmonic, $\theta_{j}(\kappa=0,z)$, switches over from monotonically
decreasing, Fig.~\ref{fig7}b, (in the situation of $U>U_{0}$, which
mimics the 3D NI/TI-overlayer heterostructure) to nonmonotonically
decreasing with minimum at $z\approx \sqrt{\mathrm{B}/|\Xi|}$,
Fig.~\ref{fig7}a, (in the situation of $U<U_{0}$, which mimics the
3D NI/NI-overlayer heterostructure). If $U\simeq U_{0}$, the EF can
be approximated as:
\begin{equation}\label{EF-U0}
\theta_{j}(0,z)\sim  \exp[-q_{1}(0,0)z] +sgn(U_{0}-U)
(\sqrt{1+|\lambda|}-\sqrt{|\lambda|})\exp[-q_{2}(0,0)z],
\end{equation}
\begin{equation}\label{q12(00)}
q_{1,2}(0,0)=\sqrt{\frac{\mathrm{B}}{|\Xi|}}(\sqrt{1+|\lambda|}\pm\sqrt{|\lambda|}).
\end{equation}
When the strength $U\rightarrow U_{\pm}$, i.e., $|E|\rightarrow
|\Xi|$, the EF (\ref{EF}) is the superposition of slow exponent
with relative small weight and quick one with relative large
weight: $q_{2}/q_{1}\simeq \sqrt{1-E^{2}/\Xi^{2}}/(2+4|\lambda|)$.
This situation is depicted with the black and red curves in
Fig.~\ref{fig7}.

\begin{figure}
 \begin{center}
\includegraphics[width=0.5\columnwidth]{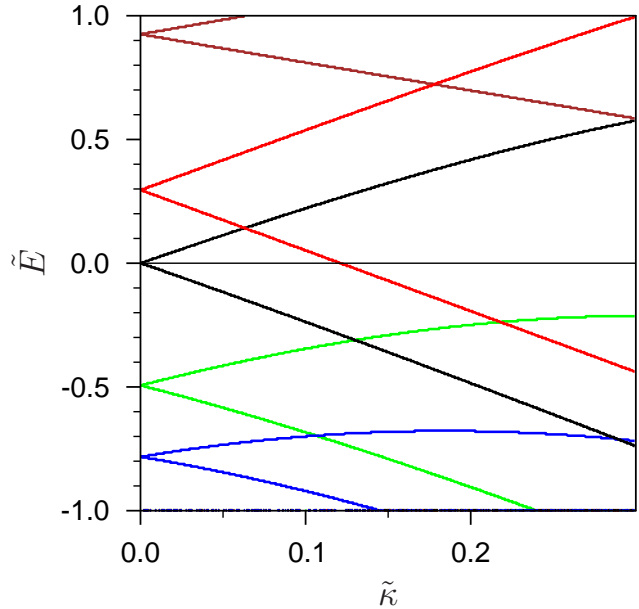}
\caption{(Color online) Spectrum of the near-surface state
$E(\kappa)$ at $U_{1}=U_{2}=U$ for several values of the SP
strength: $\tilde{U}=2.0$ (red line), $\tilde{U}=3.0$ (brown line),
$\tilde{U}=1.3$ (green line), $\tilde{U}=1.0$ (blue line),
$\tilde{U}=\sqrt{3}$ (black line), where
$\tilde{U}=\frac{dU}{\sqrt{\mathrm{B}|\Xi|}}$,
$\tilde{E}=\frac{E}{|\Xi|}$,
$\tilde{\kappa}=\sqrt{\frac{\mathrm{B}}{|\Xi|}}\kappa$,
$|\lambda|=\frac{\mathrm{A}^{2}}{4\mathrm{B}|\Xi|}=2$.}
 \label{fig8}
 \end{center}
\end{figure}

\begin{figure}
 \begin{center}
\includegraphics[width=0.5\columnwidth]{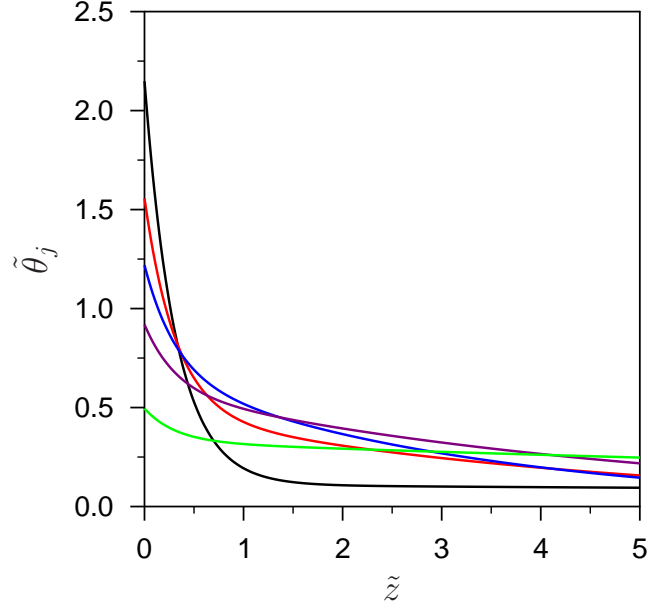}
\caption{(Color online) Space dependence of the zeroth harmonic
($\kappa=0$) of the NI envelope function $\theta_{j}(\kappa,z)$ in
the case of $U_{1}=U_{2}=U$ for several values of the SP strength:
$\tilde{U}=3.5$ (black line), $\tilde{U}=2.5$ (red line),
$\tilde{U}=1.5$ (blue line), $\tilde{U}=1.0$ (brown line),
$\tilde{U}=0.65$ (green line), where
$\tilde{U}=\frac{dU}{\sqrt{\mathrm{B}|\Xi|}}$,
$\tilde{z}=\sqrt{\frac{|\Xi|}{\mathrm{B}}}z$,
$\tilde{\theta}_{j}=\sqrt[4]{\frac{\mathrm{B}}{|\Xi|}}\theta_{j}$,
$|\lambda|=\frac{\mathrm{A}^{2}}{4\mathrm{B}|\Xi|}=2$. The envelope
function is normalized as
$\int_{0}^{\infty}dz|\theta_{j}(\bm{\kappa},z)|^{2}=1$.}
 \label{fig9}
 \end{center}
\end{figure}

Let us now draw the attention to the case  when the surface
perturbation has the spinor structure $U_{1}=U_{2}=U$ answering to
the surface electrostatic potential. Fig.~\ref{fig8} shows the
dispersion law for several values of the SP strength $U$. In the
vicinity of the points $U=\pm U_{0}=\pm \sqrt{1+|\lambda|}$, where
$|E|\ll|\Xi|$, the relations between the energy and the in-plane
momentum (in the leading order in $\kappa$) is given by
\begin{equation}\label{NI-spectrum-3}
\frac{E(\kappa)}{|\Xi|}= sgn(U)
\biggl(\frac{\tilde{U}^{2}}{\tilde{U}^{2}_{0}}-1\biggr) \pm
\frac{2|\lambda|\tilde{\kappa}}{\tilde{U}_{0}},
\end{equation}
where $\tilde{\kappa}=\kappa\sqrt{\mathrm{B}/|\Xi|}$. Thus, it is
clear, given $U$ belonging to the interval(s) $W_{-}<|U|<W_{+}$
(where $\tilde{W}_{\pm}= (\sqrt{1+4|\lambda|}\pm
\sqrt{1+2|\lambda|})/\sqrt{2}$), the surface state consists of a
single Dirac cone. Within the framework of a heuristical 'local band
bending' scheme, the variation of the strength $U$ is linked with
the relative movement of the energy levels of the 3D NI substrate
and the NI overlayer. When the SP strength value exceeds the
threshold quantity, $|U|>W_{-}$, the band structure of this system
(which consists of the two materials with a normal gap band
alignment) is inverted, i.e., either the substrate conduction band
is lower than the overlayer valence band or the substrate valence
band is higher than the overlayer conduction band. As a result, if
the strength is in the interval $W_{-}<|U|<W_{+}$, the near-surface
state of the NI covered by the normal overlayer can display a linear
dispersion dependence of the Dirac-cone form, $E(\kappa)=E(0)\pm
v\kappa$, where the node point location $E(0)$ and propagation
velocity $v$ are the functions of the strength $U$ and band
structure parameter $|\lambda|$. The energy $E(0)$ is inside the NI
bulk gap, $|E(0)|<|\Xi|$. The quasi-topological bound state is also
specified by the space distribution, which is shown in
Fig.~\ref{fig9}. The corresponding EF decays exponentially away from
the surface. When the strength $|U|$ attains the quantity $W_{\pm}$,
the near-surface state merges into the bulk continuum states, in
turn the decay length $\sim q_{2}^{-1}$ becomes large (the black and
green curves). At $|U|\simeq W_{+}$, the probability density is
concentrated close to the surface, in other case, it is rather
smeared.

\section{\emph{ab initio} calculations}

The proposed continual approach gives transparent physical
explanation for evolution of the near-surface state in both momentum
and real spaces with the SP superimposed on the TI or NI boundary.
This approach describes fairly well the electron density distribution of
the corresponding Dirac-cone-like states on the scale exceeding the
lattice spacing through EF(s) as the superposition
$\theta_{j}(\bm{\kappa},z)$ (see Eq.~(\ref{EF}), where the exponents
$q_{1,2}(\kappa,E)$
%[Eq.~(\ref{q1,2})]
and the coefficients $\alpha_{j}(\kappa,E)$, $\beta_{j}(\kappa,E)$,
are functions of the SP components and the bulk band structure
parameters. However, within the SP scheme, we are unable to
elucidate the electron density features on the scale on the order of
the SP spacing $d\ll q_{1,2}^{-1}$, in particular, capture the fine
effect of the the topological state relocation within near-surface
layers~\cite{EremeevPRB2013,Wu}. Below, in order to provide a closer
look at the wave function of the bound near-surface state and to
accurately reproduce its band structure over the whole Brillouin
zone we present \emph{ab initio} density functional theory
calculation results for some systems representing the topological
insulator substrate covered by the insulator ultrathin film.

%**************************************************************************************************************
\begin{table}
 \caption{The structural and energetic characteristics of the NI/TI Overlayer/Substrate
pairs given in the $X_1$/$X_2$ format (except for the lattice
mismatch, $\Delta$), with $X_i$ being experimental lattice
parameter, $a_i$, or calculated work function, $\Phi_i$, with the
indication of the films thicknesses, or calculated band gap, $E_i$
($i = 1, 2$). Note that band gap values, $E_i$, are given for slab
and bulk in the NI overlayer ($i=1$) and TI substrate ($i=2$) case,
respectively.}
 \label{T1}
\begin{center}
\begin{tabular}{c|cccc}
\hline
NI/TI                                              &$a_1$(\AA)/$a_2$(\AA)& $\Delta$ (\%) & $\Phi_1$(eV)/$\Phi_2$(eV) & $E_1$(eV)/$E_2$(eV)\\
\hline
$[$GeBi$_2$Te$_4]_{\mathrm {1SL}}$/Bi$_2$Te$_2$S   &  4.3225/4.316       & $+0.15$       & 4.99/5.04   & 0.44/0.27 \\
$[$Sb$_2$Te$_2$S$]_{\mathrm {1QL}}$/Sb$_2$Te$_2$Se &  4.17/4.188         & $-0.43$       & 4.99/4.62   & 0.47/0.30 \\
$[$Bi$_2$Te$_2$S$]_{\mathrm {1QL}}$/GeBi$_2$Te$_4$ &  4.316/4.3225       & $-0.15$       & 5.38/4.76   & 0.33/0.08 \\
\hline
\end{tabular}
\end{center}
\end{table}

%**************************************************************************************************************

Electronic structure calculations were carried out within the
density functional theory using the projector augmented-wave
method~\cite{Blochl1994} as implemented in the VASP
code~\cite{vasp1, vasp2}. The exchange-correlation energy was
treated using the generalized gradient approximation~\cite{PBE}.
The Hamiltonian contained the scalar relativistic corrections and
the spin-orbit coupling was taken into account by the second
variation method~\cite{Koelling.jpc1977}. In order to take into
account the effect of dispersion interactions we use the van der
Waals nonlocal correlation functional within DFT-D2
approach~\cite{Grimme.jcc2006}.

The thin film NI/TI heterostructures were simulated within a model
of repeating slabs separated by a vacuum spacing of 10~\AA. The
overlayers were symmetrically attached to both sides of the
substrate slab to preserve the inversion symmetry.

As the substrates were chosen 3D TIs with tetradymite-like layered
structures Bi$_2$Te$_2$S, Sb$_2$Te$_2$Se \cite{Menshch2011,Eremeev2}
composed of quintuple layer (QL) blocks and GeBi$_2$Te$_4$
\cite{Menshch2011_2,Eremeev1} composed of septuple layer (SL)
blocks. The substrates were simulated by 6 QL (5 SL) slabs. As the
thin insulating overlayers we used single QL(SL) films of
Bi$_2$Te$_2$S, Sb$_2$Te$_2$S, and GeBi$_2$Te$_4$ which have gapped
noninverted spectrum. The interface systems under investigation are
given in Table~\ref{T1}. As can be seen in Table~\ref{T1}, the
lattice mismatch  between overlayer and TI substrate in the
considered heterostructures doesn't exceed 0.5~\% providing very
good epitaxial compatibility. The in-plane lattice parameters of the
heterostructures were fixed to the experimental ones of the
substrate slab~\cite{Parameter_1976, Grauer_Bi2STe2,
Karpinsky1998170}. The interlayer distances within the overlayer and
the TI block, closest to the interface, were optimized.

\begin{figure}
 \includegraphics[width=\columnwidth]{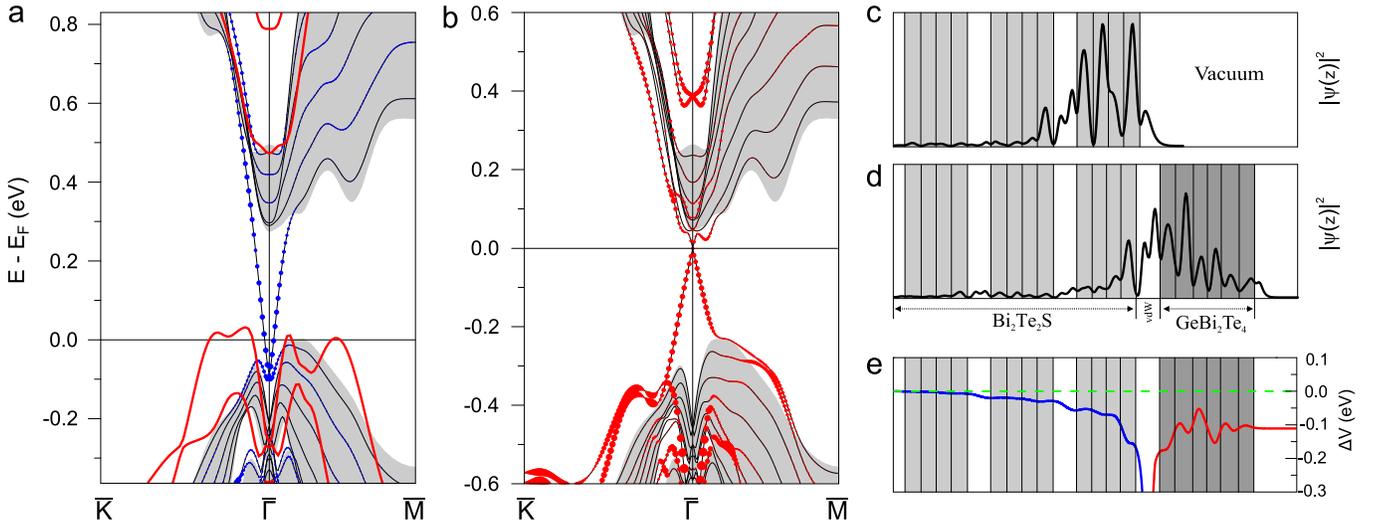}
\caption{(a) Band spectra of the  Bi$_2$Te$_2$S TI substrate and
free-standing septuble layer block of GeBi$_2$Te$_4$
%thin film
(red lines). Blue markers show the weight of the states in
near-surface QL of the substrate. Shaded areas indicate projection
of the bulk states of Bi$_2$Te$_2$S TI onto the surface Brillouin
zone; (b) Electronic spectrum of GeBi$_2$Te$_4$/Bi$_2$Te$_2$S
heterostructure. The size of red circles reflects the weight of the
states in the GeBi$_2$Te$_4$ overlayer; Spatial distribution of the
TSS charge density integrated over ($x$, $y$) plane for
Bi$_2$Te$_2$S (c) and GeBi$_2$Te$_4$/Bi$_2$Te$_2$S (d); (e) The
change in electrostatic potential $\Delta V$ of
GeBi$_2$Te$_4$/Bi$_2$Te$_2$S with respect to potentials in
free-standing Bi$_2$Te$_2$S TI substrate and GeBi$_2$Te$_4$
overlayer.}
 \label{fig10}
\end{figure}

Fig.~\ref{fig10}a shows spectra of the free-standing GeBi$_2$Te$_4$
overlayer and Bi$_2$Te$_2$S substrate. In the spectrum of
Bi$_2$Te$_2$S the topological surface state (TSS) with the Dirac
point lying in the $\bar\Gamma$ valley of the valence bulk states
propagates across the bulk energy gap. The spectrum of the
free-standing overlayer has the 440 meV gap. The energies of the
overlayer states are matched to substrate spectrum in accordance
with work functions $\Phi_1$ and $\Phi_2$ given in Tabl.~\ref{T1}.
Thus the highest occupied state of the overlayer lies $\sim 30$ meV
above the top of the valence band of substrate. The attaching of the
topologically trivial insulating overlayer to TI substrate keeps the
gapless topological surface state, however, it leads to strong
modification of the spectrum (Fig.~\ref{fig10}b): the DP shifts
towards the bottom of the conduction band of the substrate so that
above the DP it propagates as a resonant state, mixed with bulk-like
states of the Bi$_2$Te$_2$S slab. In the real space, the TSS in the
heterostructure is almost completely relocated into the overlayer
and its probability maximum lies near the
GeBi$_2$Te$_4$/Bi$_2$Te$_2$S interface plane (Fig.~\ref{fig10}c,d).
Such a behavior of the topological state can be explained by the
potential change upon the interface formation. In Fig.~\ref{fig10}e
the change in electrostatic potential of the heterostructure with
respect to potentials in free-standing substrate and overlayer is
shown. One can see that due to hybridization between orbitals of the
overlayer and substrate the potential within the TI substrate is
smoothly bent towards the interface plane while the potential within
the overlayer undergoes more noticeable changes and as a whole it
shifts down by $\sim 120$ meV with respect to its position in the
free-standing overlayer. In spite of the downward shift of the
potential the resulting topological state has the Dirac point
position higher than in the TSS of the pristine TI surface. The
change in the DP position is related to the fact that in the
heterostructure the orbitals of GeBi$_2$Te$_4$ contribute more to
the TSS than to the orbitals of Bi$_2$Te$_2$S. Thus the modification
of the topological state in the considered system is qualitatively
similar to the behavior found in the continual model with $U_1
\approx U_2$ where for negative $U_1$ beneath a critical value arise
solutions with positive shift of the Dirac point (see
Fig.~\ref{fig1}b).

\begin{figure}
 \includegraphics[width=\columnwidth]{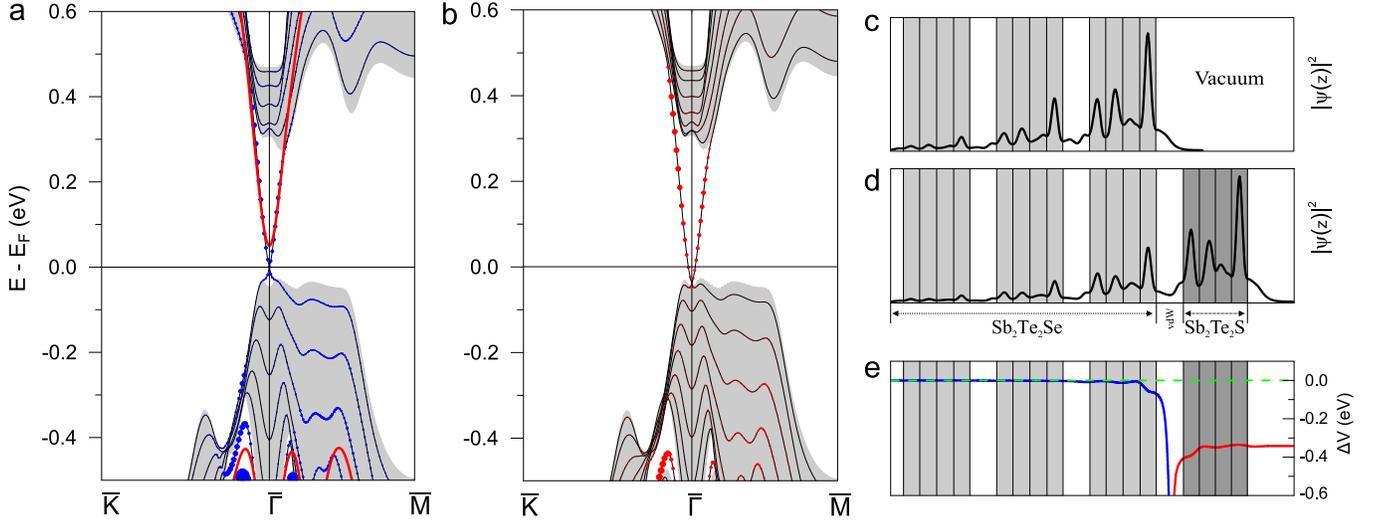}
\caption{(a) Band spectra of Sb$_2$Te$_2$Se substrate and
free-standing Sb$_2$Te$_2$S QL (red lines). Blue markers show the
weight of the states in near-surface QL of the TI substrate; (b)
Electronic spectrum of Sb$_2$Te$_2$S/Sb$_2$Te$_2$Se heterostructure;
Spatial distribution of the TSS charge density for pristine
Sb$_2$Te$_2$Se surface (c) and Sb$_2$Te$_2$S/Sb$_2$Te$_2$Se
heterostructure (d); (e) The change in electrostatic potential
$\Delta V$ of Sb$_2$Te$_2$S/Sb$_2$Te$_2$Se with respect to
potentials in free-standing Sb$_2$Te$_2$Se substrate and
Sb$_2$Te$_2$S QL.}
 \label{fig11}
\end{figure}

In the Sb$_2$Te$_2$S/Sb$_2$Te$_2$Se heterostructure
(Fig.~\ref{fig11}) the TSS dispersion remains almost unchanged with
respect to that on the pristine Sb$_2$Te$_2$Se substrate surface
being shifted towards the bulk valence band by $\sim$40~meV. Along
with this, the maximum of charge density of the topological state
relocates into the Sb$_2$Te$_2$S QL (Fig.~\ref{fig11}c,d) owing to
downward shift of the overlayer potential (Fig.~\ref{fig11}e) and
demonstrates a resonance like behavior due to the DP proximity to
the bulk continuum. In the model this scenario is realized at $U_1
\approx -U_2$ (Fig.~\ref{fig1}c) when moderate negative potential at
the interface doesn't leads to substantial change in the energy of
the TSS.

\begin{figure}
 \includegraphics[width=\columnwidth]{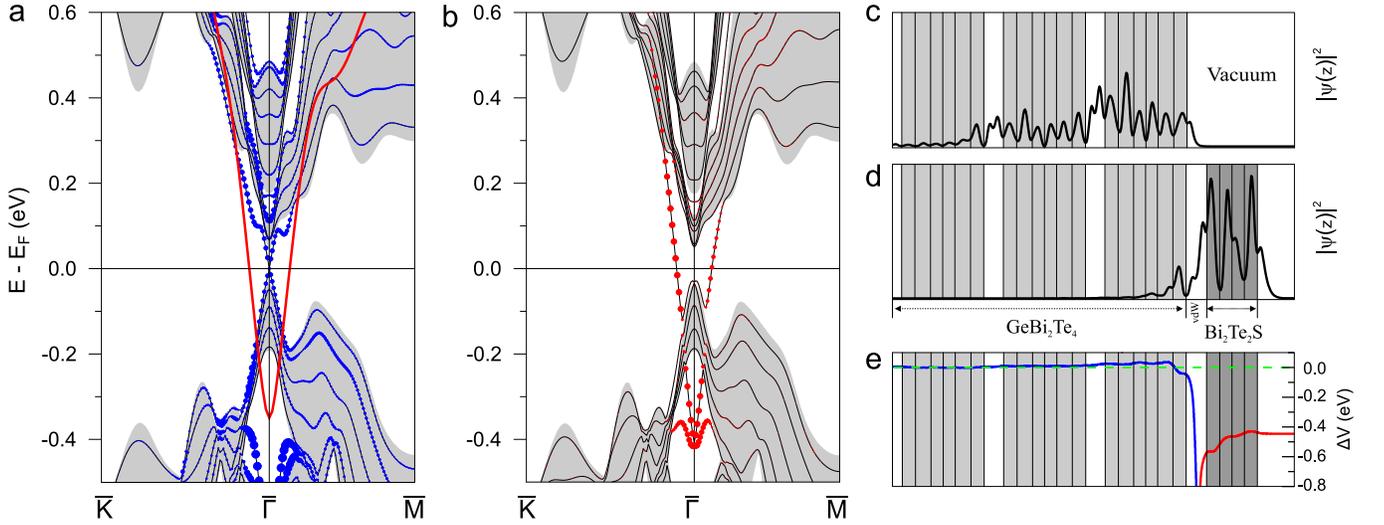}
 \caption{Electronic structure of GeBi$_2$Te$_4$ slab and free-standing
Bi$_2$Te$_2$S QL (red lines). Blue markers show the weight of the
states in near-surface SL of GeBi$_2$Te$_4$; (b) Electronic spectrum
of Bi$_2$Te$_2$S/GeBi$_2$Te$_4$ heterostructure; Spatial
distribution of the TSS charge density for pristine substrate
surface (c) and the heterostructure (d); (e) The change in
electrostatic potential $\Delta V$ of the heterostructure with
respect to potentials in free-standing GeBi$_2$Te$_4$ slab and
Bi$_2$Te$_2$S QL.}
  \label{fig12}
\end{figure}

A rather distinct type of the TSS modification demonstrate the
Bi$_2$Te$_2$S/GeBi$_2$Te$_4$ system (Fig.~\ref{fig12}). The work
function of the
%noninteracting
free standing overlayer is $0.7$ eV larger than that of the
GeBi$_2$Te$_4$ substrate slab (see Tabl.~\ref{T1}). This means that
the gap-edge states of the overlayer are far below the bulk gap of
TI substrate (Fig.~\ref{fig12}a). In contrast to the previous cases,
where at attaching of the overlayer the topological state remains
within the bulk gap of the substrate, the TSS in the
Bi$_2$Te$_2$S/GeBi$_2$Te$_4$ heterostructure occurs at -0.4 eV, in
the local $\bar\Gamma$ gap of the bulk valence band of
GeBi$_2$Te$_4$. This state lies in deep and abrupt potential well of
$\sim 500$ meV depth that causes its strong localization within
Bi$_2$Te$_2$S QL. Thus the modification of the TSS in the
Bi$_2$Te$_2$S/GeBi$_2$Te$_4$ heterostructure is qualitatively
similar to the case $U_1 \gg -U_2$ (Fig.~\ref{fig1}c) in the
continual model where large negative value of the potential results
in huge shift of the Dirac point down from its position in the
initial system.

\section{Summary and concluding remarks}

In this work, we have shown that the energy spectrum and spatial
profile of the near-surface electron states in the 3D TI
substrate/NI overlayer type of heterostructures can be controlled by
the overlayer induced SP. By choosing an appropriate overlayer it is
possible to adjust the Dirac point to the required position in the
band gap. The DFT calculations demonstrate clearly how the
overlayer-induced shift of the Dirac point from its energy position
on the pristine surface is associated with the parameters of the
band structure of the 3D TI substrate and the quintuple/septuple
overlayer. These results are in good qualitative agreement with the
tendency predicted from the analytic continual scheme, in which the
overlayer-induced change in the energy spectrum is determined by the
SP matrix elements, $U_{jj'}$. Thus the SP matrix elements can be
intuitively associated with the relative energy offsets between the
relevant band edges of the substrate and overlayer: $U_{2}\sim
\Phi_{1}-E_{1}-\Phi_{2}+E_{2}$, $U_{1}\sim \Phi_{1}-\Phi_{2}$, where
$E_{1,2}$ and $\Phi_{1,2}$ are bandgaps and work functions of an
overlayer (subscribe 1) and a TI substrate (subscribe 2).

In the framework of the continual approach, we have succeeded in
formulating general boundary conditions for the long-range EF and
finding the solution for electron bound states at the TI/NI
interface. We have obtained analytical expressions for the energy
spectrum and EF for different types and values of SP. Our results
are strictly consistent with the limiting cases of the zero and
infinite surface potentials, which were previously studied in
Refs.~\cite{Shan,Medhi}. The boundary conditions of
Eq.~(\ref{Current}) involves the SP parameters, in this sense the
represented approach has something in common with the one used to
calculate the surface states of a crystal with the relativistic band
structure~\cite{Volkov}. Note that in Ref.~\cite{Menshov}, the bound
states at the interface between 3D TI and NI have been explored on
the basis of the functional defined in the entire space. In that
work, the explicit expressions for the matrix elements of the
interface pseudo-potential, which affects electrons on the TI side
of the interface, have been analytically derived.

From the theoretical point of view, one cannot suggest universal
recipe to impose the restrictions on the EF behavior at the 3D TI
boundary for all types of the heterostructures containing 3D TIs.
Note, in particular, that the boundary conditions of
Eq.~(\ref{Current}) imposed upon the near-surface states are
different from those obtained in Ref.~\cite{Menshov}, where, for an
interface between 3D TI and a topologically trivial insulator,
authors succeeded in a formulation of the EF boundary task, the
solution of which provides insight into the electron states at the
interface. Although the EF approach may be adapted for the
description of the interface states in many semiconductor junctions,
by taking into account the general Hermiticity and symmetry
requirements~\cite{Tokatly}, this traditional description is not
always able to capture the principal features of the topological
states in the systems containing narrow-gap semiconductors with
inverted band structure and sometimes yields rather dubious
results~\cite{DeBeule}. As for the above-stated conception, the
appropriate EF boundary conditions are derived within the framework
of the formalism of the sheet-like SP. In such the approach, some
information (for example, about the effects of electron-electron
interaction) is evidently lost that is compensated by a relative
calculation simplicity and a transparent physical interpretation.
The method, conceptually presented in our work for the study of the
truncated 3D TI covered with an atomically thin non-magnetic
insulating overlayer, can be straightforwardly extended to solve a
wide range of problems related to a behavior of the near-surface
topological states under the surface perturbations listed in
Introduction.

Indeed, as it is shown in Sec.~4, the area of applicability of the
SP method significantly oversteps the formal limits of the 3D TI
substrate/NI overlayer systems. We have determined the existence
or absence of the Dirac-like near-surface modes in the
hypothetical situations when the truncated 3D NI (close to
transition into a topological phase) is brought into the contact
with the ultrathin overlayer of the 3D NI or TI material. We
predict that the near-surface ``topological-like" mode can appear
in the 3D NI substrate/NI overlayr, i.e., in the system composed
only of two topologically trivial materials. This unusual item has
something in common with the fact of the appearance of 2D TI state
in an InAs/GaSb Type-II semiconductor quantum well
(QW)~\cite{LiuC,Knez}. The unique feature of InAs/GaSb QW is that
the conduction band minimum of InAs has lower energy than the
valence band maximum of GaSb (due to the large band-offset).
Consequently, when the QW thickness is large enough, the first
electron subband of InAs layer lies below the first hole subband
of GaSb layer, i.e., an inverted band alignment, similar to that
in HgTe QWs~\cite{Hasan,Qi}, happens. The experimental study of
low temperature electronic transport have shown strong evidence
for the existence of helical edge modes in the hybridization gap
of inverted InAs/GaSb QWs~\cite{Knez}.

In view of the aforesaid, one raises the following questions
concerning different strategies to create the topological
near-surface/interface states and manage their electronic
properties: (i) How does a 3D NI overlayer influence a 3D TI? (ii)
When one puts an overlayer of one 3D TI on a substrate of another 3D
TI, what new property will come into being? (iii) How about a
hypothetical 3D TI thin film on the surface states of 3D NI? (iiii)
Finally, one could ask how to construct a quasi-topological bound
state at the boundary between two topologically trivial insulators?
Our work partly answers these questions. The approach proposed above
unveils the physics of the near-surface states in the semiconductor
heterostructures containing 3D TIs. At the same time, it might be
considered as a good guidebook for the qualitative interpretation
and forecast of the topological phase behavior tendencies in 3D TIs
under surface perturbations. The obtained results should open new
opportunities to design various combinations of topological and
conventional materials for electronic/spintronic applications.

\newpage

%%%%%%%%%%%%%%%%%%%%%%%%%%%%%%%%%%%%%%%%%%%%%%%%%%%%%%%%%%%%%%%%%%%%%%%%%%%%%%%%%%%%%%%%%%%%%%%%%%%%%%%%%%%%%
%%%%%%%%%%%%%%%%%%%%%%%%%%%%%%%%%%%%%%%%%%%%%%%%%%%%%%%%%%%%%%%%%%%%%%%%%%%%%%%%%%%%%%%%%%%%%%%%%%%%%%%%%%%%%

%%%%%%%%%%%%%%%%%%%%%%%%%%%%%%%%%%%%%%%%%%%%%%%%%%%%%%%%%%%%%%%%%%%%%%%%%%%%%%%%%%%%%%%%%%%%%%%%%%%%%%%%%%%%%
%%%%%%%%%%%%%%%%%%%%%%%%%%%%%%%%%%%%%%%%%%%%%%%%%%%%%%%%%%%%%%%%%%%%%%%%%%%%%%%%%%%%%%%%%%%%%%%%%%%%%%%%%%%%%

%%%%%%%%%%%%%%%%%%%%%%%%%%%%%%%%%%%%%%%%%%%%%%%%%%%%%%%%%%%%%%%%%%%%%%%%%%%%%%%%%%%%%%%%%%%%%%%%%%%%%%%%%%%%%
%%%%%%%%%%%%%%%%%%%%%%%%%%%%%%%%%%%%%%%%%%%%%%%%%%%%%%%%%%%%%%%%%%%%%%%%%%%%%%%%%%%%%%%%%%%%%%%%%%%%%%%%%%%%%

%%%%%%%%%%%%%%%%%%%%%%%%%%%%%%%%%%%%%%%%%%%%%%%%%%%%%%%%%%%%%%%%%%%%%%%%%%%%%%%%%%%%%%%%%%%%%%%%%%%%%%%%%%%%%
%%%%%%%%%%%%%%%%%%%%%%%%%%%%%%%%%%%%%%%%%%%%%%%%%%%%%%%%%%%%%%%%%%%%%%%%%%%%%%%%%%%%%%%%%%%%%%%%%%%%%%%%%%%%%

%%%%%%%%%%%%%%%%%%%%%%%%%%%%%%%%%%%%%%%%%%%%%%%%%%%%%%%%%%%%%%%%%%%%%%%%%%%%%%%%%%%%%%%%%%%%%%%%%%%%%%%%%%%%%
%%%%%%%%%%%%%%%%%%%%%%%%%%%%%%%%%%%%%%%%%%%%%%%%%%%%%%%%%%%%%%%%%%%%%%%%%%%%%%%%%%%%%%%%%%%%%%%%%%%%%%%%%%%%%

%%%%%%%%%%%%%%%%%%%%%%%%%%%%%%%%%%%%%%%%%%%%%%%%%%%%%%%%%%%%%%%%%%%%%%%%%%%%%%%%%%%%%%%%%%%%%%%%%%%%%%%%%%%%%
%%%%%%%%%%%%%%%%%%%%%%%%%%%%%%%%%%%%%%%%%%%%%%%%%%%%%%%%%%%%%%%%%%%%%%%%%%%%%%%%%%%%%%%%%%%%%%%%%%%%%%%%%%%%%

%%%%%%%%%%%%%%%%%%%%%%%%%%%%%%%%%%%%%%%%%%%%%%%%%%%%%%%%%%%%%%%%%%%%%%%%%%%%%%%%%%%%%%%%%%%%%%%%%%%%%%%%%%%%%
%%%%%%%%%%%%%%%%%%%%%%%%%%%%%%%%%%%%%%%%%%%%%%%%%%%%%%%%%%%%%%%%%%%%%%%%%%%%%%%%%%%%%%%%%%%%%%%%%%%%%%%%%%%%%

%%%%%%%%%%%%%%%%%%%%%%%%%%%%%%%%%%%%%%%%%%%%%%%%%%%%%%%%%%%%%%%%%%%%%%%%%%%%%%%%%%%%%%%%%%%%%%%%%%%%%%%%%%%%%
%%%%%%%%%%%%%%%%%%%%%%%%%%%%%%%%%%%%%%%%%%%%%%%%%%%%%%%%%%%%%%%%%%%%%%%%%%%%%%%%%%%%%%%%%%%%%%%%%%%%%%%%%%%%%

%%%%%%%%%%%%%%%%%%%%%%%%%%%%%%%%%%%%%%%%%%%%%%%%%%%%%%%%%%%%%%%%%%%%%%%%%%%%%%%%%%%%%%%%%%%%%%%%%%%%%%%%%%%%%
%%%%%%%%%%%%%%%%%%%%%%%%%%%%%%%%%%%%%%%%%%%%%%%%%%%%%%%%%%%%%%%%%%%%%%%%%%%%%%%%%%%%%%%%%%%%%%%%%%%%%%%%%%%%%

%%%%%%%%%%%%%%%%%%%%%%%%%%%%%%%%%%%%%%%%%%%%%%%%%%%%%%%%%%%%%%%%%%%%%%%%%%%%%%%%%%%%%%%%%%%%%%%%%%%%%%%%%%%%%
%%%%%%%%%%%%%%%%%%%%%%%%%%%%%%%%%%%%%%%%%%%%%%%%%%%%%%%%%%%%%%%%%%%%%%%%%%%%%%%%%%%%%%%%%%%%%%%%%%%%%%%%%%%%%

%%%%%%%%%%%%%%%%%%%%%%%%%%%%%%%%%%%%%%%%%%%%%%%%%%%%%%%%%%%%%%%%%%%%%%%%%%%%%%%%%%%%%%%%%%%%%%%%%%%%%%%%%%%%%
%%%%%%%%%%%%%%%%%%%%%%%%%%%%%%%%%%%%%%%%%%%%%%%%%%%%%%%%%%%%%%%%%%%%%%%%%%%%%%%%%%%%%%%%%%%%%%%%%%%%%%%%%%%%%

%%%%%%%%%%%%%%%%%%%%%%%%%%%%%%%%%%%%%%%%%%%%%%%%%%%%%%%%%%%%%%%%%%%%%%%%%%%%%%%%%%%%%%%%%%%%%%%%%%%%%%%%%%%%%
%%%%%%%%%%%%%%%%%%%%%%%%%%%%%%%%%%%%%%%%%%%%%%%%%%%%%%%%%%%%%%%%%%%%%%%%%%%%%%%%%%%%%%%%%%%%%%%%%%%%%%%%%%%%%

\end{document}